\documentclass[sigconf]{acmart}

\AtBeginDocument{%
  \providecommand\BibTeX{{%
    \normalfont B\kern-0.5em{\scshape i\kern-0.25em b}\kern-0.8em\TeX}}}

\copyrightyear{2023} 
\acmYear{2023} 
\setcopyright{acmlicensed}\acmConference[MM '23]{Proceedings of the 31st ACM International Conference on Multimedia}{October 29-November 3, 2023}{Ottawa, ON, Canada}
\acmBooktitle{Proceedings of the 31st ACM International Conference on Multimedia (MM '23), October 29-November 3, 2023, Ottawa, ON, Canada}
\acmPrice{15.00}
\acmDOI{10.1145/3581783.3611723}
\acmISBN{979-8-4007-0108-5/23/10}

\usepackage{algorithm}
\usepackage{algorithmicx}
\usepackage{algpseudocode}
\usepackage{multirow}
\usepackage{balance}
\usepackage{subfigure}
\usepackage{enumitem}

\usepackage{ragged2e}

\newcommand{\nosection}[1]{\vspace{2pt}\noindent\textbf{#1.}}

\newcommand{\modelname}{\textbf{PBAT}}

\begin{document}

\title{Personalized Behavior-Aware Transformer for Multi-Behavior Sequential Recommendation}

\author{Jiajie Su}
\affiliation{College of Computer Science
\institution{Zhejiang University,China}
\country{}
}
\email{sujiajie@zju.edu.cn}

\author{Chaochao Chen}
\affiliation{College of Computer Science
\institution{Zhejiang University,China}
\country{}
}
\email{zjuccc@zju.edu.cn}

\author{Zibin Lin}
\affiliation{College of Computer Science
\institution{Zhejiang University,China}
\country{}
}
\email{zibinlin@zju.edu.cn}

\author{Xi Li}
\affiliation{College of Computer Science
\institution{Zhejiang University,China}
\country{}
}
\email{xilizju@zju.edu.cn}

\author{Weiming Liu}
\affiliation{College of Computer Science
\institution{Zhejiang University,China}
\country{}
}
\email{21831010@zju.edu.cn}

\author{Xiaolin Zheng}
\authornote{Xiaolin Zheng is the corresponding author.}
\affiliation{College of Computer Science
\institution{Zhejiang University,China}
\country{}
}
\email{xlzheng@zju.edu.cn}

\renewcommand{\shortauthors}{Jiajie Su et al.}


\begin{abstract}
Sequential Recommendation (SR) captures users' dynamic preferences by modeling how users transit among items.
However, SR models that utilize only single type of behavior interaction data encounter performance degradation when the sequences are short.
To tackle this problem, we focus on Multi-Behavior Sequential Recommendation (MBSR) in this paper, which aims to leverage time-evolving heterogeneous behavioral dependencies for better exploring users' potential intents on the target behavior.
Solving MBSR is challenging.
On the one hand, users exhibit diverse multi-behavior patterns due to personal characteristics.
On the other hand, there exists comprehensive co-influence between behavior correlations and item collaborations, the intensity of which is deeply affected by temporal factors.
To tackle these challenges, we propose a \underline{P}ersonalized \underline{B}ehavior-\underline{A}ware \underline{T}ransformer framework (PBAT) for MBSR problem, which models personalized patterns and multifaceted sequential collaborations in a novel way to boost recommendation performance.
%
%
First, PBAT develops a personalized behavior pattern generator in the representation layer, which extracts dynamic and discriminative behavior patterns for sequential learning.
Second, PBAT reforms the self-attention layer with a behavior-aware collaboration extractor, which introduces a fused behavior-aware attention mechanism for incorporating both behavioral and temporal impacts into collaborative transitions.
We conduct experiments on three benchmark datasets and the results demonstrate the effectiveness and interpretability of our framework.
Our implementation code is released at https://github.com/TiliaceaeSU/PBAT.
\end{abstract}


\maketitle

\section{Introduction}

Recommender systems have been widely employed on online platforms for providing accurate content to mitigate information explosion.
Sequential Recommendation (SR) \cite{kang2018self,tang2018personalized,wang2020next,zheng2022ddghm} is one of the core tasks in recommender systems, which aims at capturing users' time-varying interests in terms of their historical behaviors and make next-item recommendation.
%

With the advance in neural networks, various SR models \cite{hidasi2015session,sun2019bert4rec,wu2019session,xie2020contrastive} have effectively extracted sequential signals for inferring purchase intentions.
However, most of these methods are designed specially for single-behavior setting, which makes them face the difficulty in characterizing user preferences when single-behavior sequences are short, i.e., the data sparsity problem.
Actually, except for the most common target behavior \textit{purchase}, there are plenty of other interaction forms on an E-commerce platform in practice such as \textit{click}, \textit{like}, and \textit{add-to-cart}, which reflect more fine-grained and meaningful user preferences that are hidden underneath sequential context.
To address above issues, in this paper, we focus on the Multi-Behavior Sequential Recommendation (MBSR), 
which seeks to capture and leverage dynamic collaborative signals between \textit{auxiliary behaviors} and the \textit{target behavior} in multi-behavior sequences towards better recommendation results.


Some pioneering studies have investigated on MBSR problem.
%
%
A series of researches \cite{dmt,wang2020beyond,yu2022graph} divide multi-behavior sequences by behavior types, extracting co-influence among behavior-specific sub-sequences.
Another branch of work \cite{li2018learning,zhou2018micro,tanjim2020attentive} models item and behavior sequences independently, considering behavior knowledge as an auxiliary component of representations.
Recent studies \cite{yuan2022multi,yang2022multi} inject behavior interaction context into item transitions, exploring behavior-aware transitional patterns of user interests.
Although these methods reach the breakthrough of exploiting sequential patterns in the behavioral perspective, the MBSR problem remains non-trivial due to the following challenges:
\begin{figure}[t]
 \begin{center}
 \includegraphics[width=\columnwidth]{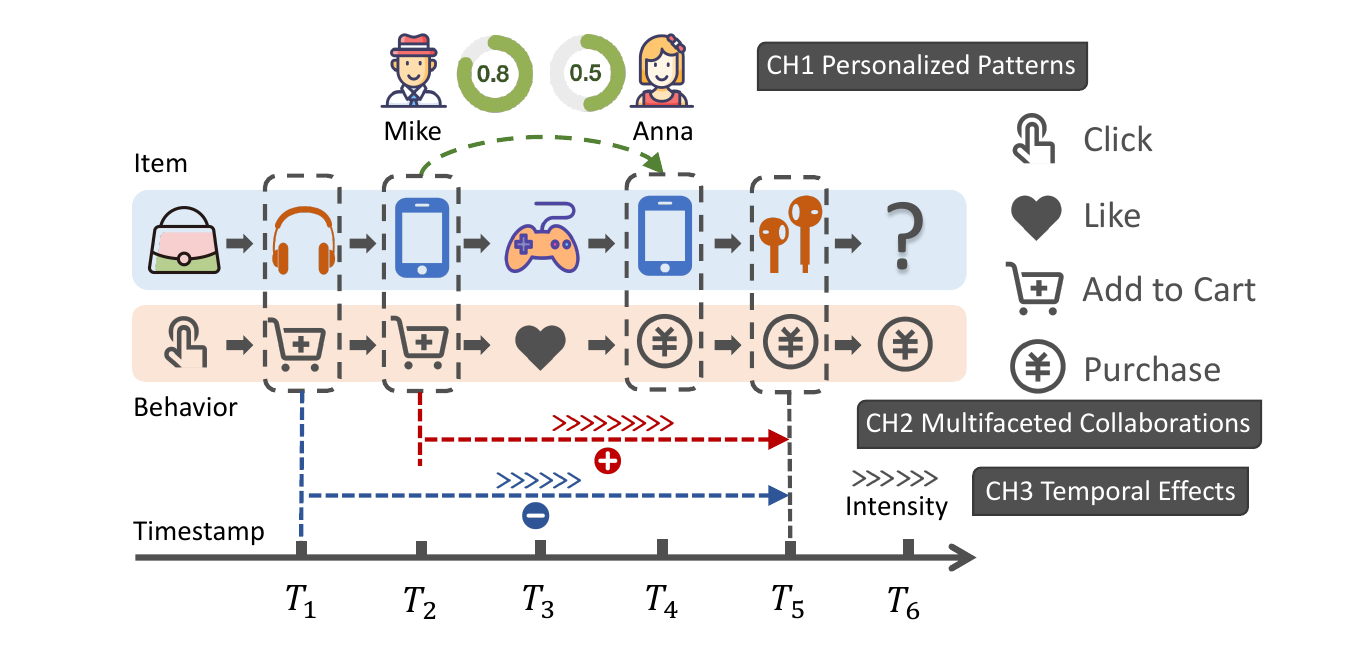}
 \vspace{-0.25 in}
 \caption{A motivating example.
 }
 \vspace{-0.25 in} 
 \label{fig:story}
 \end{center}
\end{figure}
\begin{itemize}[leftmargin=*]
    \item \textbf{Ch1. Personalized Multi-Behavior Patterns.}
    %
    Different users exhibit various online behavioral habits due to their personal characteristics.
    As Figure \ref{fig:story} illustrates, Mike used to add items, he was highly likely to buy, to his shopping cart. 
    But for Anna, there is no strong correlation between adding to cart and an actual purchase.
    Inter-dependencies between different types of interaction behaviors vary by users.
    Therefore, modeling sequential behavioral patterns with a fixed paradigm for all users is not appropriate, and how to depict personalized multi-behavior patterns remains challenging.

    \item \textbf{Ch2. Multifaceted Behavior-Aware Collaborations.}
    Interplay within the same type of behavior pairs presents diverse collaborations among different items.
    In Figure \ref{fig:story}, take the behavior pair \textit{Cart-to-Purchase} as an example.
    Adding a cellphone into cart drives user to buy Bluetooth earphones later since they are complementary in usage.
    While adding headphones into cart reduces the probability of purchasing earphones because of product competitiveness.
    Behavior sequences imply potential collaborative item relations, and conversely item collaborations influence how behavior transitions make an impact.
    Thus, mutual influence across item-side and behavior-side sequential dependencies makes it difficult to exploit multifaceted collaborations.

    \item \textbf{Ch3. Temporal Effects on Collaboration Intensity.}
    The position distance between an interaction pair directly affects the degree of their correlation.
    Intuitively, the smaller the interval is, the stronger the collaboration shows.
    Many previous studies \cite{sun2019bert4rec,yang2022multi} take position information into consideration when modeling sequential patterns.
    But they usually encode position tags into item representations, and few of them investigates how temporal intervals directly influences behavior-aware collaborations.
    Exploring complex ternary relationships in sequential context among item, behavior, and position is a remaining problem.

\end{itemize}


To address aforementioned issues, we propose \underline{P}ersonalized \underline{B}ehavior-\underline{A}ware \underline{T}ransformer (PBAT) for solving MBSR problem.
The core insight of \modelname~ is leveraging multi-behavioral interaction information to promote exploring users' dynamic and discriminative latent intentions for next-item prediction on the target behavior.
%
To achieve that, we design two modules, i.e., \textit{personalized behavior pattern generator} and \textit{behavior-aware collaboration extractor}.
The \textit{personalized behavior pattern generator} module is developed to exploit diverse behavior patterns that are hidden underneath multi-behavior sequences with respect to user personalities (\textbf{Ch1}).
In this module, we first introduce a \textit{dynamic representation encoding} method which utilizes Gaussian distributions to describe all entities and relations in the MBSR scenario, i.e., user, item, behavior, position, and behavior-relation, so as to precisely capture the dynamics and uncertainty of sequential representations.
Then, in order to generate behavior patterns with more personalization, we employ \textit{personalized pattern learning} that proposes Self-Adaptive Gaussian Production (SAGP) for organically integrating unified behavior features with personal interests.
In this way, sequential behavioral pattern learning facilitates the pertinence in user-level and improves the robustness in sequence-level.
The \textit{behavior-aware collaboration extractor} module is proposed to realize two goals: i) mine multifaceted behavior-aware collaborations across both item-side and behavior-side (\textbf{Ch2}), and 
ii) measure how time intervals affect collaboration intensity (\textbf{Ch3}).
To this end, we attentively merge behavior correlations and personalized patterns to extract the \textit{behavioral collaboration impact factor} as the first step.
After that, we propose a \textit{fused behavior-aware attention} mechanism, equipped with the TriSAGP method, to depict ternary
associations in position-enhanced multi-behavior collaborations as the second step. 

The main contributions are summarized as follows:
(1) We propose a framework which introduces a new paradigm to reveal the nature of personalized behavior patterns for improving multi-behavior sequential recommendation performance.
(2) We design a personalized behavior pattern generator which uncovers distinguishing 
multi-behavior patterns to better portray individualized preferences.
(3) We develop a behavior-aware collaboration extractor which investigates multifaceted collaborations under both behavioral and temporal impacts for accurately exploring sequential dependencies. 
(4) We conduct experiments on three benchmark datasets and the
results demonstrate the effectiveness of our model.


\section{Related Work}

\nosection{Sequential Recommendation}
SR aims to characterize evolving user preferences by modeling sequences. 
Early work usually models the sequential dependencies with the Markov Chain assumption \cite{rendle2010factorizing}. 
With the advance in deep learning, Recurrent Neural Networks (RNN) based \cite{hidasi2015session,hidasi2018recurrent,liu2023joint,wu2017recurrent}, Convolutional Neural Networks (CNN) based \cite{tang2018personalized}, Graph Neural Networks (GNN) based \cite{wu2019session,zheng2020dgtn,su2023enhancing}, and attention based \cite{kang2018self,sun2019bert4rec,wu2020sse,li2022coarse} models have been adopted to explore dynamic user interests that are hidden underneath behavior sequences.
Recently, contrastive learning based models \cite{xie2020contrastive,qiu2021memory,lin2022dual} are introduced to extract meaningful user patterns by deriving self-supervision signals.
While these methods promote sequential recommendation performance, they suffer from weak prediction power when single-behavior sequences are short.

\nosection{Multi-Behavior Recommendation}
Multi-Behavior Recommendation (MBR) is built to utilize multiple user-item interactions for enhancing recommendation on target behaviors.
Firstly, early studies \cite{singh2008relational,krohn2012multi,zhao2015improving} on MBR are mainly based on matrix factorization, which simultaneously factorize multiple user-item interaction matrices with sharing item or user embeddings.
Later, some deep learning methods show superiority in modeling multi-behavior data \cite{chen2020efficient,nmtr,chen2021graph}, such as NMTR \cite{nmtr} that combines neural collaborative filtering and multitask learning to exploit behavior patterns.
Motivated by strengths of GNNs, more researches \cite{mbgcn,zhang2020multiplex,mbgmn} employ multi-relational interaction graphs for MBR.
%
Recently, some advanced models adopt contrastive learning \cite{yang2021hyper,wei2022contrastive} or variational autoencoder \cite{ma2022vae++} and achieve promising performance as well.
Different from these studies, we study MBSR problem in this paper, which introduces sequential information into MBR and requires more accurate portraying on behavior-aware sequential patterns.

\nosection{Multi-Behavior Sequential Recommendation}
Existing studies on MBSR have two main types with respect to modeling paradigm.
\textit{First}, a series of work \cite{dmt,dipn,mgnn,chen2021curriculum} divide multi-behavior sequences into sub-sequences according to specific behavior interaction types, which extracts sequential features in each sub-sequence and synthesizes them jointly.
%
%
For instance, DMT \cite{dmt} models users’ different types of behavior sequences simultaneously with multiple transformers.
MGNN-SPred \cite{mgnn} builds a multi-relational item graph based on behavior-specific sub-sequences to explore global item-to-item relations.
This type of methods effectively extracts behavior-specific sequential patterns but neglects multi-behavior collaborations in sequential context.
\textit{Second}, another branch of work \cite{zhou2018micro,meng2020incorporating,binn,tanjim2020attentive,yuan2022multi,yang2022multi} separates item and behavior sequential modeling phases.
They utilize behavior patterns as auxiliary information through adding behavior types into input or modeling behavior sequences independently.
For example, BINN \cite{binn} employs a contextural long short-term memory architecture to model item and behavior sequences.
%
MB-STR \cite{yuan2022multi} exploits both behavior-specific semantics and multi-behavior sequential heterogeneous dependencies via transformer layers.
This type of methods retains integrity of interaction sequences which enables exploration of complex multi-behavior sequential patterns.
In this paper, our framework further develops the second type of MBSR, which specifically learns personalized behavior patterns and explores multifaceted collaborations under impacts from multiple factors.

\section{METHODOLOGY}

\subsection{Problem Formulation}
We formulate a typical MBSR scenario.
Let $\mathcal{U}=\{u_1,u_2,\ldots,u_{|\mathcal{U}|}\}$ and $\mathcal{V}=\{v_1,v_2,\ldots,v_{|\mathcal{V}|}\}$ represent the set of users and items, with $|\mathcal{U}|$ and $|\mathcal{V}|$ denoting the size of each set.
Similarly, the behavior set is written as $\mathcal{B}=\{b_1,b_2,\ldots,b_{|\mathcal{B}|}\}$, where $|\mathcal{B}|$ denotes the amount of behavior types.
We define \textit{purchase} as the \textbf{target behavior} for prediction and other behavior types as \textbf{auxiliary behaviors}.
Then a \textbf{multi-behavior interaction sequence} is formulated as $\mathcal{S}_{u}=\left\{\left<v^u_1, b^u_1\right>,\ldots,\left<v^u_i, b^u_i\right>,\ldots,\left<v^u_L, b^u_L\right>\right\}$, where $\left<v^u_i, b^u_i\right>$ is the $i$-th behavior-aware interaction pair, describing that item $v^u_i \in \mathcal{V}$ is interacted by user $u \in \mathcal{U}$ with behavior $b^u_i \in \mathcal{B}$, and $L$ indicates the sequence length.
The goal of MBSR is exploring multi-behavior sequential dependencies in $\mathcal{S}_u$ and predicting top-$K$ items from $\mathcal{V}$ that are most likely to be interacted by user $u$ under target behavior at the next timestamp.
%
%
The notation table is in Appendix A.

\subsection{An Overview of \modelname}

\begin{figure*}
\centering
\includegraphics[width=1\textwidth]{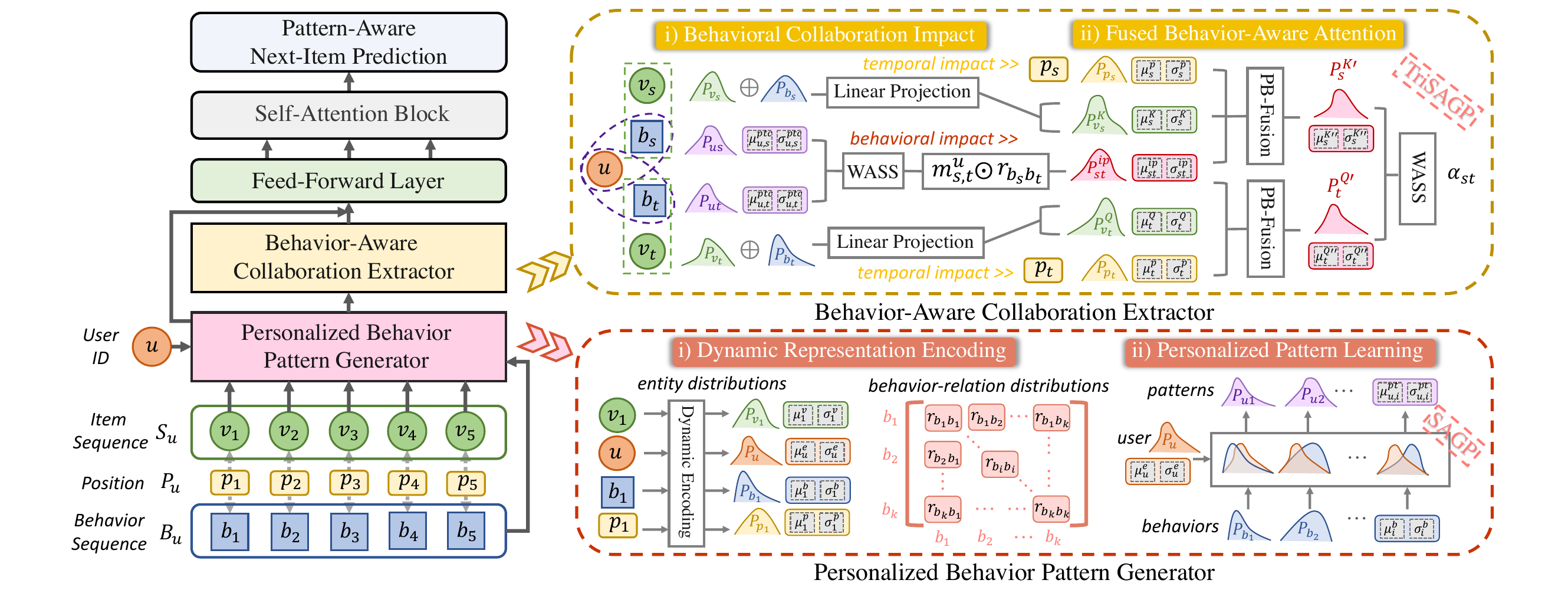}
\vspace{-0.8cm}
\caption{The framework of \modelname. We present the detailed zoom-in view of two main modules, i.e., personalized behavior pattern generator and behavior-aware collaboration extractor, in the right panel.
}
\label{fig:framework}
\vspace{-0.4cm} 
\end{figure*}

The aim of \modelname~is to provide better sequential recommendation performance by modeling dynamic multi-behavior patterns.
We present the overall framework of our model in Figure \ref{fig:framework}.
\modelname~is a transformer-based SR, which mainly consists of two modules, i.e., (1) personalized behavior pattern generator and (2) behavior-aware collaboration extractor.
Firstly, to address the challenge about portraying diverse personalized multi-behavior patterns, we design the \textbf{personalized behavior pattern generator}.
The generation lays in two folds: a) \textit{dynamic representation encoding} that utilizes Gaussian distributions to depict more discriminative entities and relations in multi-behavior sequences, and 
b) \textit{personalized pattern learning} which leverages self-adaptive Gaussian production to refine general behavior pattern with user customs.
Secondly, to overcome the challenges of exploiting multifaceted behavior-aware collaborations and capturing temporal interval effects on collaboration intensity, we propose the \textbf{behavior-aware collaboration extractor}.
This module is realized in two steps: a) extract behavioral collaboration impact factors by integrating unified behavior relations with personalized patterns.
b) apply a fused behavior-aware attention mechanism which comprehensively explores complex sequential collaborations from item, behavior, and position perspectives.
Finally, the model completes the pattern-aware next-item prediction.

\subsection{Personalized Behavior Pattern Generator}
Traditional transformer-based SR methods effectively model sequential dependencies of item transitions, but encounter obstacles when extracting user interests in a behavior-aware manner.
Behavior transitions express complicated semantics on explaining how users make their decisions, resulting the failure of the unbiased learning paradigm.
To this end, we introduce \textbf{personalized behavior pattern generator} which is composed of two parts, i.e., \textit{dynamic representation encoding} and \textit{personalized pattern learning}, to capture diverse behavioral characteristics of different users.

\subsubsection{Dynamic Representation Encoding}
In this part, we introduce how to encode item and behavior transitions with distribution embedding to provide more generalized representations.
Affected by both external and internal factors, user behavior patterns show a large extent of dynamicity and uncertainty in the sequential context.
Thus, fixed vectors are insufficient to describe evolving sequential patterns.
According to \cite{zheng2019deep,fan2022sequential}, distribution representation learning produces uncertainties and endows embeddings with more flexibility.
Rather than deterministic embedding, we adopt multi-dimensional elliptical Gaussian distributions to deal with uncertain features of MBSR entities, i.e., user, item, position, and behavior.

\nosection{Entity distribution}
A typical elliptical Gaussian distribution is represented by a mean vector and a covariance vector, where the mean shows the general distinguishing features and the covariance controls the uncertainty.
For all items, we initialize a mean embedding table as $\textbf{H}^\mu=[\boldsymbol{h}_1^\mu,\boldsymbol{h}_2^\mu,...,\boldsymbol{h}_{|\mathcal{V}|}^\mu]$ and a covariance embedding table as $\textbf{H}^\sigma=[\boldsymbol{h}_1^\sigma,\boldsymbol{h}_2^\sigma,...,\boldsymbol{h}_{|\mathcal{V}|}^\sigma]$, where $\textbf{H}^\mu,\textbf{H}^\sigma \in \mathbb{R}^{|\mathcal{V}| \times D}$.
$D$ indicates the dimension of representations.
Then items embeddings in sequences can be formulated as:
\begin{equation}
\begin{aligned}
    \textbf{M}^\mu_{\mathcal{S}_u} = [\boldsymbol{h}^\mu_{v^u_1},\boldsymbol{h}^\mu_{v^u_2},\ldots,\boldsymbol{h}^\mu_{v^u_L}], \quad
    \textbf{M}^\sigma_{\mathcal{S}_u} = [\boldsymbol{h}^\sigma_{v^u_1},\boldsymbol{h}^\sigma_{v^u_2},\ldots,\boldsymbol{h}^\sigma_{v^u_L}].
\end{aligned}
\nonumber
\end{equation}
For example, the item $v_1^u$ is represented as $d$-dimensional Gaussian distribution $\mathcal{N}(\mu^{v}_{u,1},\sigma^{v}_{u,1})$, where $\mu^{v}_{u,1}=\boldsymbol{h}^\mu_{v^u_1}$ and $\sigma^{v}_{u,1}=diag(\boldsymbol{h}^\sigma_{v^u_1})$.
Similarly, we initialized embeddings as $\textbf{E}^\mu=[\boldsymbol{e}_1^\mu,\boldsymbol{e}_2^\mu,\ldots,\boldsymbol{e}_{|\mathcal{U}|}^\mu]$ and $\textbf{E}^\sigma=[\boldsymbol{e}_1^\sigma,\boldsymbol{e}_2^\sigma,\ldots,\boldsymbol{e}_{|\mathcal{U}|}^\sigma]$ for users, $\textbf{B}^\mu=[\boldsymbol{b}_1^\mu,\boldsymbol{b}_2^\mu,\ldots,\boldsymbol{b}_{|\mathcal{B}|}^\mu]$ and $\textbf{B}^\sigma=[\boldsymbol{b}_1^\sigma,\boldsymbol{b}_2^\sigma,\ldots,\boldsymbol{b}_{|\mathcal{B}|}^\sigma]$ for behaviors, respectively.
Since item position affects sequential learning, we set  learnable position distribution embeddings $\textbf{P}^\mu=[\boldsymbol{p}_1^\mu,\boldsymbol{p}_2^\mu,\ldots,\boldsymbol{p}_{L}^\mu],\textbf{P}^\sigma=[\boldsymbol{p}_1^\sigma,\boldsymbol{p}_2^\sigma,\ldots,\boldsymbol{p}_{L}^\sigma]$ as well.
%

\nosection{Behavior-relation distribution}
Due to heterogeneous sequential dependencies that each pair of behavior transition exhibits, we additionally introduce behavior-relation distribution.
Each relation between a pair of behavior types is regarded as an independent distribution representation, e.g., the relation between $b_i$ and $b_j$ is governed by its unique mean $\boldsymbol{r}^\mu_{b_ib_j}$ and covariance vector $\boldsymbol{r}^\sigma_{b_ib_j}$.
Then we obtain behavior-relation distribution table $\textbf{R}=[\textbf{R}_\mu,\textbf{R}_\sigma]$:

\begin{equation}
    \textbf{R}_\mu\left(i,j\right)= \boldsymbol{r}^\mu_{b_ib_j},\ \textbf{R}_\sigma\left(i,j\right)= \boldsymbol{r}^\sigma_{b_ib_j},\ 1\leq i,j \leq K,
\nonumber
\end{equation}
where $\textbf{R}_\mu, \textbf{R}_\sigma \in \mathbb{R}^{K \times K \times D}$, and $K$ denotes the amount of interaction behavior types in the dataset.

\subsubsection{Personalized Pattern Learning}
We then present how to exploit personalized behavior patterns with dynamic distribution representation. 
Intuitively, the general behavior mode expresses various variants on the individual, which makes user-behavior pattern learning necessary.
To address the issue, we develop a Self-Adaptive Gaussian Production (SAGP) to combine the unified behavior transition features with personal characteristics.

Recall that both users and behaviors are presented by distribution embeddings with mean vector determining global features and covariance vector referring to uncertainty.
Then we define the personalized pattern for user $u$ on $i$-th type of behavior:
\begin{equation}
\left[\mu^{pt}_{u,i},\sigma^{pt}_{u,i}\right]=SAGP(\left[\mu^e_u,\sigma^e_u\right],[\mu^b_i,\sigma^b_i]).
\nonumber
\end{equation}
Here the superscript $e$, $b$, and $pt$ demote user entity, behavior entity, and pattern entity respectively.
Note that user $u$ is represented by $\mathcal{N}(\mu^e_u,\sigma^e_u)$, behavior $i$ is represented by $\mathcal{N}(\mu^b_i,\sigma^b_i)$.
The design principle of SAGP is to integrate major features from both user and behavior perspectives and restrict the uncertainty range as well.
%
First, we formulate the mean vector that implys feature fusion of personalized behavior patterns:
\begin{equation}
    \mu^{pt}_{u,i} = 
    \alpha_1\mu^e_u+\alpha_2\textbf{W}^{b}_\mu\mu^b_i.
\nonumber
\end{equation}
Here, we set $\alpha_1 = \frac{{\sigma^b_i}^2}{{\sigma^b_i}^2+{\sigma^e_u}^2}$ and $\alpha_2 = \frac{{\sigma^e_u}^2}{{\sigma^b_i}^2+{\sigma^e_u}^2}$, which are parameters that balance the influence from user and behavior terms.
Since there exists distribution bias between different feature spaces, i.e., user space and behavior space, we utilize the projection weight $\textbf{W}^{b}_\mu$ for feature alignment.
To obtain robust behavior patterns for promoting sequential learning, we tend to extract discriminative features with strong stability from original user and behavior entity.
Since lower covariances indicate more accurate distributions, a relatively large covariance should bring less effect from the entity to the final pattern.
Second, the covariance vector is generated by
\begin{equation}
\small
    {\sigma^{pt}_{u,i}}^2 = \frac{2{\sigma^e_u}^2{\sigma^b_i}^2}{{\sigma^e_u}^2+{\sigma^b_i}^2},
\nonumber
\end{equation}
where $\sigma^{pt}_{u,i}$ is bounded by $\left[\operatorname{Min}(\sigma^e_u, \sigma^b_i),\operatorname{Max}(\sigma^e_u,\sigma^b_i)\right]$, so that the uncertainty of learned patterns is further restrained.

\subsection{Behavior-Aware Collaboration Extractor}
The heterogeneity of behavior patterns hiddened underneath interaction sequences not only exists in different individuals, but also shows across multiple item collaborations.
Owing to the intrinsic structure of multi-behavior sequences, collaborative relations within item and behavior transitions are intertwined, making it a great challenge to portray the bi-directional influence.
To solve this problem, we propose a \textbf{behavior-aware collaboration extractor} to replace ordinary attention layers in the traditional transformer.
The extraction is done in two steps, i.e., produce \textit{behavioral collaboration impact factor} and realize \textit{fused behavior-aware attention}.
We present the algorithm description of this module in Appendix B.

\subsubsection{Behavioral Collaboration Impact Factor}
To capture behavior transition semantics in sequential context, we apply an Wasserstein-based method to measure behavioral collaboration impact factors.

Formally, given two items $v^u_s$ and $v^u_t$ at the $s$-th and $t$-th positions in the sequence of user $u$, the corresponding behaviors are $b_s$ and $b_t$, respectively.
Through SAGP, we have gotten two personalized patterns as $\mathcal{N}(\mu^{pt}_{u,s},\sigma^{pt}_{u,s})$ and $\mathcal{N}(\mu^{pt}_{u,t},\sigma^{pt}_{u,t})$.
To begin with, we apply a linear projection on the patterns:
\begin{equation}
    \mu^{ptc}_{u,x}=\mu^{pt}_{u,x}\textbf{W}^c_{\mu}, \quad
\sigma^{ptc}_{u,x}=\operatorname{\textbf{ELU}}\left(\sigma^{pt}_{u,x}\textbf{W}^c_{\sigma}\right)+1,
\nonumber
\end{equation}
where $\textbf{ELU}$ denotes exponential linear unit that maps inputs into $[-1, +\infty]$ to guarantee the positive definite property of covariance \cite{fan2022sequential}, and $x \in \{s,t\}$.
Then we adopt Wasserstein distance to measure behavioral collaborative impacts between personalized patterns:
\begin{equation}
\begin{aligned}
    m^u_{s,t} &= Wass\left(\left[\mu^{ptc}_{u,s},\sigma^{ptc}_{u,s}\right],\left[\mu^{ptc}_{u,t},\sigma^{ptc}_{u,t}\right]\right),
\nonumber
\end{aligned}
\end{equation}
where Wasserstein distance is calculated as:
\begin{equation}
\small
    Wass_{12} = \Vert \mu_{1} -\mu_{2} \Vert_2^2 + \textbf{Tr}\left(\sigma_{1}+\sigma_{2}-2{\left({\sigma_{1}}^{1/2}{\sigma_{1}}{\sigma_{2}}^{1/2}\right)}^{1/2}\right).
\nonumber
\end{equation}
The reason of using Wasserstein distance lays in twofold: (1) Wasserstein distance provides a measurement between two distributions with uncertainty information, and meanwhile brings a stable and robust training process \cite{kolouri2018sliced}. 
(2) It is proved to satisfy triangle inequality for probability measures on separable metric spaces, which guarantees the collaborative transitivity in behavioral sequential pattern learning \cite{clement2008elementary}.
After getting the degree coefficient $m^u_{s,t}$ that indicates co-influence between patterns especially for user $u$, we pick out the corresponding behavior-relation representation from relation table $\textbf{R}$, and attentively combine the general behavior inter-correlation features with personalized patterns:
\begin{equation}
    \mu^{ip}_{s,t}= m^u_{s,t}\boldsymbol{r}^\mu_{b_sb_t},\quad
    \sigma^{ip}_{s,t}= m^u_{s,t}\boldsymbol{r}^\sigma_{b_sb_t}.
\nonumber
\end{equation}
Here, we use $\textit{ip}$ to denote the final impact factor, which is represented by a Gaussian distribution $\mathcal{N}(\mu^{ip}_{s,t},\sigma^{ip}_{s,t})$.

\subsubsection{Fused Behavior-Aware Attention}
Sequential dependency in multi-behavior sequences is the outcome under mutual effects from behavior correlations and item transitions.
In this part, we introduce a fused behavior-aware attention mechanism, which has two purposes: (1) extracting the comprehensive sequential collaborations fused from item- and behavior-side.
(2) modeling temporal effects brought by position information on behavior-aware collaborations.

\nosection{Position-enhanced behavior-aware fusion}
To realize multi-head self-attention, we perform linear projection on the incorporation of item and behavior representations to get behavior-specific queries, keys, and values for each head. 
Taking item $v_s$ as example: 
\begin{equation}
\begin{aligned}
    \mu^Q_s &= \mu^v_{s}\textbf{W}^{QI}_{\mu}+\mu^b_{s}\textbf{W}^{QB}_{\mu}, \quad
\sigma^Q_s=\operatorname{\textbf{ELU}}\left(\sigma^v_{s}\textbf{W}^{QI}_{\sigma}+\sigma^b_{s}\textbf{W}^{QB}_{\sigma}\right)+1 ,\\
    \mu^K_s &= \mu^v_{s}\textbf{W}^{KI}_{\mu}+\mu^b_{s}\textbf{W}^{KB}_{\mu}, \quad
\sigma^K_s=\operatorname{\textbf{ELU}}\left(\sigma^v_{s}\textbf{W}^{KI}_{\sigma}+\sigma^b_{s}\textbf{W}^{KB}_\sigma\right)+1 ,\\
\mu^V_s &= \mu^v_{s}\textbf{W}^{VI}_{\mu}+\mu^b_{s}\textbf{W}^{VB}_{\mu}, \quad 
\sigma^V_s=\operatorname{\textbf{ELU}}\left(\sigma^v_{s}\textbf{W}^{VI}_{\sigma}+\sigma^b_{s}\textbf{W}^{VB}_\sigma\right)+1,
\end{aligned}
\nonumber
\end{equation}
where the query, key, and value are distributions as $\mathcal{N}(\mu^Q_s,\sigma^Q_s)$, $\mathcal{N}(\mu^K_s,\sigma^K_s)$, and $\mathcal{N}(\mu^V_s,\sigma^V_s)$.
Note that we omit the head subscript here for clarity.
For traditional attention layer, dot-product is usually applied to measure the correlation between items, but it is not appropriate for inferring the distance between Gaussian distributions.
In this case, we propose a Position-enhanced Behavior-aware Fusion (PB-Fusion) which is specially designed for mixture distribution representations, to measure the discrepancy between multi-behavior interaction pairs.
%
%
Apart from incorporating sequential features from item-side and behavior-side, we further consider temporal impacts on collaborations as well.
To do that, we propose TriSAGP method that extends SAGP from duality to ternary association, injecting corresponding position information into the fusion stage.
Formally, for the item pair $v_s$ and $v_t$, we merge their behavioral impact factor $\mathcal{N}(\mu^{ip}_{s,t},\sigma^{ip}_{s,t})$ and temporal impact factor $\mathcal{N}(\mu^{p}_{s},\sigma^{p}_{s})$ into queries and keys through TriSAGP as:
\begin{equation}
\small
\begin{aligned}
&\left[\mu^{K'}_s,\sigma^{K'}_s\right]=TriSAGP\left(\left[\mu^K_s,\sigma^K_s\right],\left[\mu^{ip}_{s,t},\sigma^{ip}_{s,t}\right],\left[\mu^{p}_{s},\sigma^{p}_{s}\right]\right),\\
    &{\sigma^{K'}_s}^2 = \left(\frac{1}{{\sigma^K_s}^2}+\frac{1}{{\sigma^{ip}_{s,t}}^2}+\frac{1}{{\sigma^p_s}^2}\right)^{-1},
    \mu^{K'}_s = {\sigma^{K'}_s}^2 \cdot\left(\frac{\mu^K_s}{{\sigma^K_s}^2}+\textbf{W}^{ip}_\mu\frac{\mu^{ip}_{s,t}}{{\sigma^{ip}_{s,t}}^2}+\textbf{W}^{p}_\mu\frac{\mu^{p}_{s}}{{\sigma^{p}_{s}}^2}\right).
\end{aligned}
\nonumber
\end{equation}
Here, we denote the new distribution of the query as $\mathcal{N}\left(\mu^{K'}_s,\sigma^{K'}_s\right)$.
$\textbf{W}^{ip}_\mu$ and $\textbf{W}^{p}_\mu$ are projection weights for feature space alignment.
Similarly, the new key can be obtained as $\mathcal{N}\left(\mu^{Q'}_t,\sigma^{Q'}_t\right)$.
%

\nosection{Attentive aggregation}
With behavior-aware keys and queries, we calculate the attention score by Wasserstein distance as $\alpha_{s,t}=-Wass\left(\left[ \mu^{K'}_{s}, \sigma^{K'}_{s}\right],\left[ \mu^{Q'}_{t}, \sigma^{Q'}_{t}\right]\right)$
Then we normalize the attention scores and utilize them to aggregate the weighted sum of each self-attention layer.
The aggregations of mean and covariance obey the linear combination property of Gaussian distribution\cite{dwyer1958generalizations}:
\begin{equation}
    x^\mu_t = \sum^L_{j=1}\frac{\alpha_{j,t}}{\sum^L_{i=1}\alpha_{i,t}}\mu^V_j, \quad
    x^\sigma_t = \sum^L_{j=1}{\left(\frac{\alpha_{j,t}}{\sum^L_{i=1}\alpha_{i,t}}\right)}^2\sigma^V_j.
\nonumber
\end{equation}
We denote outputs of the attention layer from head $i$ as  $X^i_\mu=\left(x_1^\mu,x_2^\mu,\ldots,x_L^\mu\right), X^i_\sigma=\left(x_1^\sigma,x_2^\sigma,\ldots,x_L^\sigma\right)$.
With integrating $h$ heads, we have $X_\mu=\operatorname{Concat}\left(X^1_\mu,\ldots,X^h_\mu\right)$ and $X_\sigma=\operatorname{Concat}\left(X^1_\sigma,\ldots,X^h_\sigma\right)$.

\nosection{Feed-forward layer and self-attention block}
As previous studies \cite{kang2018self,yuan2022multi}, we design feed-forward layers (FFL) and self-attention blocks (SAB) to obtain complex context information in multi-behavior sequences.
We present the details of design formulas of this part in Appendix C.
Finally, the outputs of mean and covariance embeddings $X_\mu^{(n)}$ and $X_\sigma^{(n)}$ are generated by the $n$-th block.

\subsection{Prediction and Training}

\nosection{Pattern-aware next-item prediction}
After hierarchically and attentively extracting sequential dependencies of previous interacted items, we predict the next-item under the target behavior.
To enhance the personalization in the prediction phase, we refine the behavior-aware sequential representation $\mathcal{N}(x^\mu_t,x^\sigma_t)$ at the $t$-th position in the sequence through integrating user-specific behavior pattern $\mathcal{N}(\mu^{pt}_{u,z},\sigma^{pt}_{u,z})$ as $\left[\hat{x}^\mu_t,\hat{x}^\sigma_t\right]=SAGP\left(\left[x^\mu_t,x^\sigma_t\right],\left[\mu^{pt}_{u,z},\sigma^{pt}_{u,z}\right]\right),
\nonumber$
Here, $z$ indicates the target behavior type, and $\mathcal{N}(\hat{x}^\mu_t,\hat{x}^\sigma_t)$ denotes the refined state.
Then we compute recommendation scores for each target item $v_i$ at $(t+1)$-th position by Wasserstein distance:
\begin{equation}
\hat{\boldsymbol{y}}_{v_i,t+1} = -Wass\left(\left[\hat{x}^\mu_t,\hat{x}^\sigma_t\right],\left[\mu^v_i,\sigma^v_i\right]\right).
\nonumber
\end{equation}

\nosection{Optimization and training}
To efficiently train our model, we adopt the well-known \textit{Cloze task} \cite{sun2019bert4rec} as the training objective.
In this multi-behavior sequential case, for each training step, we randomly mask $\rho$ proportion of items in the sequence, i.e., replace the item with special token $[mask]$ but keep the corresponding behavior token unmasked.
Then our model makes prediction on masked items based on sequential context and the target behavior pattern.
To measure the ranking prediction loss, we apply cross-entropy loss as \cite{kang2018self,wu2020sse}:
\begin{equation}
\small
    \mathcal{L}_{pre} = \sum_{\mathcal{S}_u\in\mathcal{S}}\sum_{t\in \mathcal{P}_m}\sum_{v_j\notin \mathcal{S}_u}-\left[\log\left(\sigma\left(\hat{\boldsymbol{y}}_{v_t,t}\right)\right)+\log\left(1-\sigma\left(\hat{\boldsymbol{y}}_{v_j,t}\right)\right)\right],
\nonumber
\end{equation}
where $v_j$ denotes a uniformly sampled negative item, $\mathcal{P}_m$ denotes the position set of masked items in $\mathcal{S}_u$, and $\sigma$ is the sigmoid function.
Note that we pad all sequences to the same length $L$ and ignore the loss value when $v_t$ is a padding item.

\section{EXPERIMENTS AND ANALYSIS}


In this section, we conduct experiments on several public datasets to answer the following questions: 
(1) \textbf{RQ1}: How does our model perform compared
with the state-of-the-art traditional recommendation, SR, MBR, and MBSR methods
(2) \textbf{RQ2}: How does each designed module of our model contribute to the final performance?
(3) \textbf{RQ3}: How do different types of auxiliary  behaviors contribute to the prediction of target behavior?
(4) \textbf{RQ4}: How to prove that \modelname~extracts personalized behavior patterns and multifaceted collaborations in an interpretable way?
(4) \textbf{RQ5}: How does the performance of \modelname~vary with different values of hyper-parameters?

\subsection{Experimental Setup} 
\nosection{Dataset}
We evaluate our model on three public benchmark datasets: \textbf{Taobao}, \textbf{IJCAI}, \textbf{Yelp} \cite{matn,mbgmn,yuan2022multi}. 
Particularly, \textbf{Taobao} comes from one of the largest e-commerce platforms named Taobao, which contains four types of behaviors, i.e., \textit{page view}, \textit{tag-as-favorite}, \textit{add-to-cart}, and \textit{purchase}.
\textbf{IJCAI} is released by IJCAI competition for user activity modeling from an online business-to-consumer e-commerce, which also includes four types of behaviors, i.e., \textit{page view}, \textit{tag-as-favorite}, \textit{add-to-cart}, and \textit{purchase}.
\textbf{Yelp} is a widely used benchmark dataset that comes from Yelp.
Following \cite{xia2020multiplex,yuan2022multi}, we differentiate the interaction data into three types of behavior according to the rating, i.e., \textit{like}, \textit{neutral}, and \textit{dislike}.
%
%
Besides, an additional \textit{tip} behavior is included, which indicates that users write tips on visited venues.
Note that we set the target behavior as \textit{purchase} for dataset Taobao and IJCAI, and \textit{like} for Yelp.
The detailed statistics of datasets are summarized in Appendix D.

    


 
\nosection{Evaluation Protocols}
Following \cite{kang2018self,sun2019bert4rec}, we adopt the leave-one-out evaluation strategy that regards the last two interactions as validation and test data, respectively, and those before the penultimate interaction as training data.
We choose two evaluation metrics, i.e., Hit Rate (HR) and Normalized Discounted Cumulative
Gain (NDCG), where we set the
cut-off of ranked lists as 5 and 10.
For all experiments,
we repeat them five times and report average results.

\begin{table*}[t]
\renewcommand\arraystretch{0.7}
\centering
\caption{Experimental results on three datasets. The best results are boldfaced and the second-best results are underlined.}
\vspace{-0.4cm}
\label{tab:compare}
\resizebox{\linewidth}{!}{
\begin{tabular}{ccccccccccccc}
\toprule
\multirow{1}{*}{}
& \multicolumn{4}{c}{\textbf{Taobao}}
& \multicolumn{4}{c}{\textbf{IJCAI}}
& \multicolumn{4}{c}{\textbf{Yelp}}\\
\cmidrule(lr){2-5} \cmidrule(lr){6-9} \cmidrule(lr){10-13}
&HR@5 &NDCG@5 &HR@10 &NDCG@10 &HR@5 &NDCG@5 &HR@10 &NDCG@10 &HR@5 &NDCG@5 &HR@10 &NDCG@10 \\
\midrule
DMF & 0.189	&0.135	&0.298	&0.185	&0.275	&0.197	&0.379	&0.242	&0.655	&0.415	&0.743	&0.474\\
NGCF & 0.196	&0.130	&0.318	&0.197	&0.346	&0.232	&0.463	&0.293	&0.674	&0.428	&0.778	&0.495\\

\midrule 
GRU4REC &0.241	&0.176	&0.368	&0.215	&0.498	&0.347	&0.592	&0.391	&0.684	&0.432	&0.781	&0.495\\
Caser &0.204 &0.132	&0.341	&0.193	&0.329	&0.213	&0.439	&0.274	&0.673	&0.425	&0.763	&0.489 \\
SASRec &0.253	&0.182	&0.377	&0.226	&0.489	&0.351	&0.604	&0.412	&0.728	&0.451	&0.793	&0.501\\
BERT4Rec &0.278	&0.193	&0.396	&0.241	&0.493	&0.362	&0.622	&0.426	&0.745	&0.473	&0.824	&0.528\\

\midrule 
NMTR &0.183	&0.159	&0.328	&0.168	&0.396	&0.269	&0.495	&0.312	&0.694	&0.445	&0.785	&0.467\\
MB-GCN &0.247	&0.172	&0.362	&0.215	&0.345	&0.209	&0.453	&0.269	&0.734	&0.468	&0.807	&0.513\\
MB-GMN &0.394	&0.221	&0.498	&0.304	&0.467	&0.286	&0.524	&0.337	&0.736	&0.525	&0.862	&0.580\\

\midrule 

DIPN &0.256	&0.157	&0.332	&0.184	&0.425	&0.259	&0.481	&0.305	&0.557	&0.452	&0.683	&0.517\\
BINN &0.358	&0.224	&0.463	&0.306	&0.398	&0.242	&0.469	&0.291	&0.375	&0.344	&0.489	&0.413\\
MGNN-SPred &0.353	&0.218	&0.459	&0.304	&0.381	&0.234	&0.443	&0.279	&0.372	&0.335	&0.485	&0.402\\
DMT &0.547	&0.355	&0.654	&0.408	&0.592	&0.354	&0.675	&0.458	&0.549	&0.471	&0.662	&0.525\\
ASLI &0.382	&0.255	&0.503	&0.325	&0.436	&0.278	&0.502	&0.314	&0.417	&0.392	&0.528	&0.457\\
MBHT &0.687	&\underline{0.590} &0.764 &\underline{0.615} &0.774 & 0.675 &0.852 &0.701 &0.727 &0.550 &0.861 &0.594\\
MB-STR & \underline{0.691}	&0.585	&\underline{0.772}	&0.611	&\underline{0.805}	&\underline{0.692}	&\underline{0.883}	&\underline{0.717}	&\underline{0.750}	&\underline{0.574}	&\underline{0.874}	&\underline{0.615}\\

\midrule 
\modelname &\textbf{0.735}	&\textbf{0.651}	&\textbf{0.802}	&\textbf{0.673}  & \textbf{0.871}	&\textbf{0.781}	&\textbf{0.924}	&\textbf{0.798} & \textbf{0.758} & \textbf{0.584} & \textbf{0.889} &\textbf{0.627}\\
Impv. & 6.37\% & 10.3\% & 3.89\% & 9.43\%& 8.20\%& 12.9\% & 4.64\% & 11.3\% &1.07\% & 1.74\% & 1.72\% & 1.95\%\\

\bottomrule

\end{tabular}
}
\vspace{-0.10 in}
\end{table*}

\nosection{Parameter settings}
We implement the proposed model using Pytorch.
We use Gaussian distribution $\mathcal{N}(0,0.02)$ to initialize entity and behavior-relation distributions in \modelname~and choose the Adam optimizer with a learning rate of 0.001 for optimization.
For a fair comparison, we tune the parameters of \modelname~and baseline models to their best values.
We set the training batch size as 128 and max sequence length $L$ as 50, which are consistent with \cite{yuan2022multi}.
We utilize the similar structure of BERT4REC \cite{sun2019bert4rec} as the transformer backbone, where we adopt the multi-head mechanism with head number as 2 per layer.
And the mask proportion $\rho$ of multi-behavior sequences is empirically set to 0.2.
Specifically, we study the effect of the embedding dimension $D$ by varying it in $\{8,16,32,64,128\}$, and the effect of attention block number $N$ by varying it in $\{1,2,3,4\}$.

\subsection{Baseline Algorithms}

\nosection{Traditional Recommendation}
%
(1) \textbf{DMF} \cite{xue2017deep} utilizes matrix factorization to learn low-dimensional factors of users/items.  
%
%
(2) \textbf{NGCF} \cite{wang2019neural} explores collaborative signals in user-item integration graphs.

\nosection{Sequential Recommendation}
(3) \textbf{GRU4Rec} \cite{hidasi2015session} utilizes gate recurrent units to encode sequential information.
(4) \textbf{Caser} \cite{tang2018personalized} uses CNNs to capture both general preferences and sequential dependencies.
(5) \textbf{SASRec} \cite{kang2018self} is a classical transformer-based model that adaptively learns sequential patterns.
(6) \textbf{BERT4Rec} \cite{sun2019bert4rec} adopts a bi-directional transformer to model user sequences.

\nosection{Multi-behavior Recommendation}
(7) \textbf{NMTR} \cite{nmtr} develops a multi-task learning paradigm to model different behavior correlations in a cascaded manner.
(8) \textbf{MB-GCN} \cite{mbgcn} proposes GCN to re-construct multiple user-item interactions through behavior-aware message passing.
(9) \textbf{MB-GMN} \cite{mbgmn} integrates multi-behavior pattern modeling into meta-learning in a graph meta network.

\nosection{Multi-behavior Sequential Recommendation}
(10) \textbf{DIPN} \cite{dipn} extracts user intents by leveraging bi-directional recurrent network and hierarchical attention mechanism.
(11) \textbf{BINN} \cite{binn} converts sequential items into a unified space and discriminatively learns behavior information on LSTM-based architectures.
(12) \textbf{MGNN-SPred} \cite{mgnn} constructs multi-relational graphs with behavior-specific sub-sequences and integrates representations by gating mechanisms.
(13) \textbf{DMT} \cite{dmt} models multi-behavior sequences with multifaceted transformers and utilizes mixture-of-experts to optimize objectives.
(14) \textbf{ASLI} \cite{tanjim2020attentive} learns item similarities via self-attention layers and explore user intents through a convolutional network. 
%
%
(15) \textbf{MBHT} \cite{yang2022multi} proposes a multi-scale transformer enhanced with hypergraphs to encode behavior-aware sequential patterns.
(16) \textbf{MB-STR} \cite{yuan2022multi} is a recent transformer-based model that captures both behavior-specific semantics and multi-behavior dependencies.

\subsection{Model Comparison (RQ1)}
We evaluate performance of next-item prediction on the target behavior among \modelname~and all baselines.
We report comparison results on three datasets in Table \ref{tab:compare}.
From the results, we can conclude: 
(1) \textbf{Capturing sequential dependencies benefits recommendation performance.}
    In general, SR baselines outperform most of traditional ones that do not utilize sequential information.
    We can draw similar conclusion by comparing MBSR models with MBR baselines.
    %
(2) \textbf{Leveraging multi-behavior information further promotes user preference modeling.}
    Most MBR models show their superiority when compared with traditional models.
    %
    %
    Compared with SR methods, lots of MBSR methods also boost the prediction performance.
    %
(3) \textbf{\modelname~shows its effectiveness for MBSR problem.}
    Our approach \modelname~shows superiority over all type of  baselines in terms of all metrics.
    The average improvement of \modelname~to the best baseline is $6.37\%$ and $10.3\%$ for HR@5 and NDCG@5 on Taobao, $8.20\%$ and $12.9\%$ on IJCAI, $1.07\%$ and $1.74\%$ on Yelp.
    %
    %
    First, compared with RNN-based (DIPN, BINN) and GNN-based (MGNN-SPred) models, \modelname~adopts the transformer backbone, which takes full advantage of self-attention mechanism to extract and integrate both long- and short-term sequential dependencies.
    Second, \modelname~achieves the best performance among transformer-based (DMT, ASLI, MBHT, and MB-STR) models, which can be attributed from:
    i) With personalized behavior pattern generator, \modelname~depicts dynamic and discriminative personal multi-behavior patterns that play an important role when mining users' potential preferences.
    ii) Through behavior-aware collaboration extractor, \modelname~captures multifaceted collaborations across item-side and behavior-side under temporal impacts, which further enhances multi-relational sequential learning.

\begin{table}[t]
\renewcommand\arraystretch{0.7}
\centering
\caption{Ablation study on key components of \modelname. Note that the cut-off of the ranked list is 10.}
\vspace{-0.4cm}
\label{tab:ablation}
\resizebox{\linewidth}{!}{
\begin{tabular}{ccccccc}
\toprule
\multirow{1}{*}{}
& \multicolumn{2}{c}{\textbf{Taobao}}
& \multicolumn{2}{c}{\textbf{IJCAI}}
& \multicolumn{2}{c}{\textbf{Yelp}}\\
\cmidrule(lr){2-3} \cmidrule(lr){4-5} \cmidrule(lr){6-7}
&HR &NDCG &HR &NDCG &HR &NDCG\\
\midrule
\textit{w/o \textbf{PBPG}} & 0.755 & 0.628 & 0.853 &0.739 & 0.844 & 0.572 \\
\textit{w/o \textbf{BACE}} & 0.672 & 0.533 & 0.715 &0.572 & 0.658 & 0.513 \\

\midrule 
\textit{w/o \textbf{fob}} & 0.715 & 0.577 & 0.801	&0.673 & 0.746 & 0.551\\
\textit{w/o \textbf{fop}} & 0.763 & 0.639 & 0.865	& 0.752 & 0.861 & 0.595 \\
\textit{w/o \textbf{fobp}}& 0.687 & 0.548 & 0.744	& 0.609 & 0.667 & 0.532\\

\midrule 
\modelname & \textbf{0.802} & \textbf{0.673} & \textbf{0.924} & \textbf{0.798} &\textbf{0.889} &\textbf{0.627}\\
\bottomrule

\end{tabular}
}
\vspace{-0.15 in}
\end{table}

\begin{figure} 
\centering
        \subfigure[Taobao-HR]{\label{fig:subfig:a}
\resizebox{0.42\hsize}{!}{\includegraphics[width=0.47\linewidth]{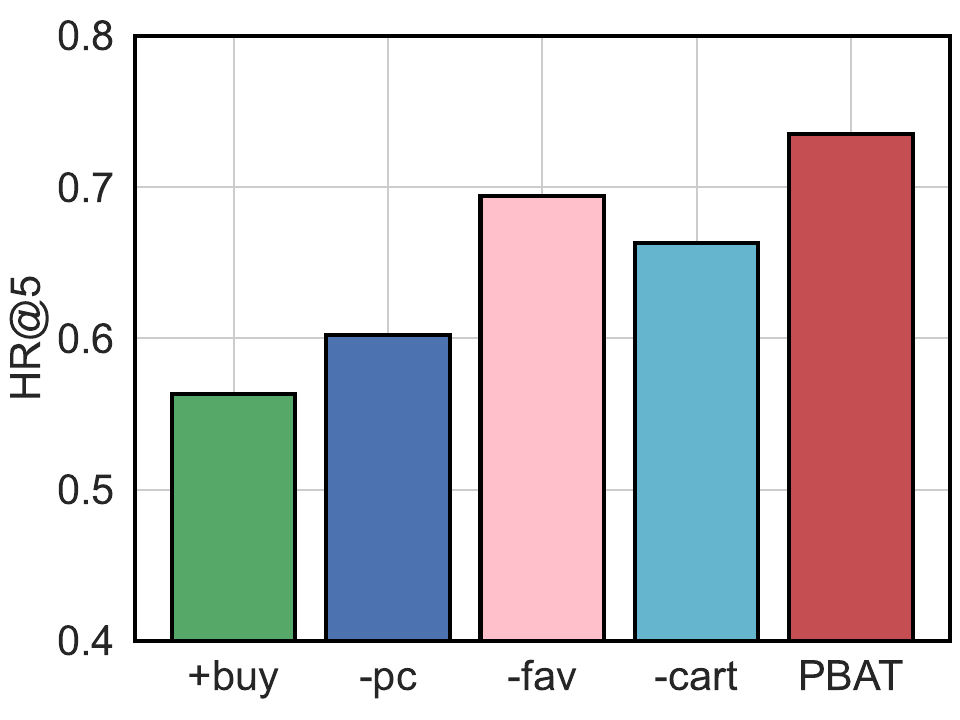}}}
\hspace{0.01\linewidth}
\subfigure[Taobao-NDCG]{\label{fig:subfig:b}
\resizebox{0.42\hsize}{!}
{\includegraphics[width=0.47\linewidth]{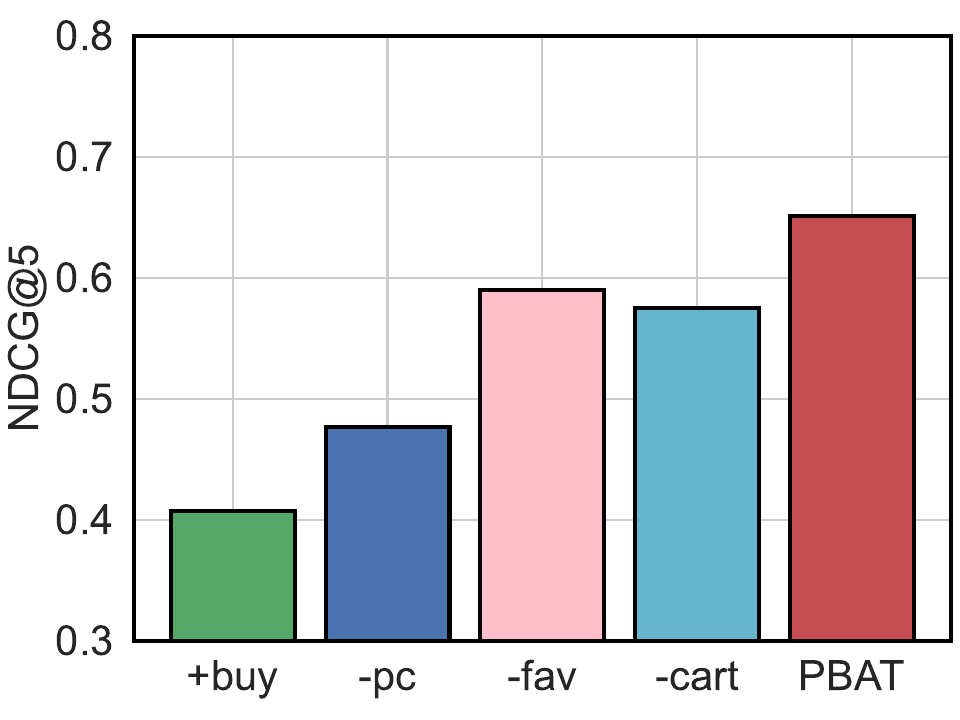}}}\\
\subfigure[IJCAI-HR]{\label{fig:subfig:c}
\resizebox{0.42\hsize}{!}
{\includegraphics[width=0.47\linewidth]{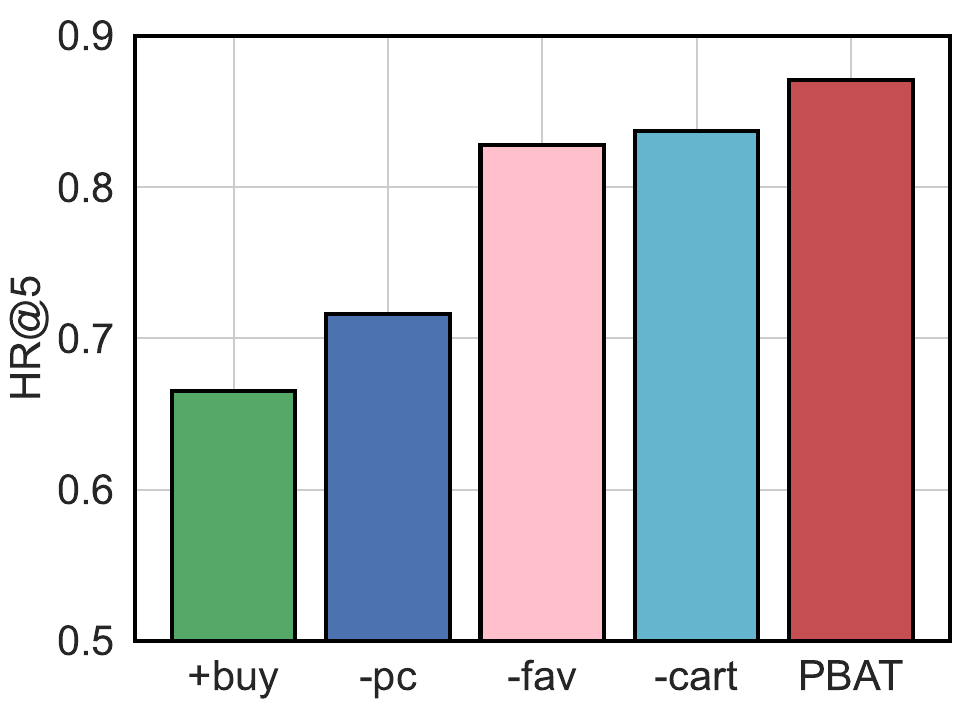}}}
\hspace{0.01\linewidth}
\subfigure[IJCAI-NDCG]{\label{fig:subfig:d}
\resizebox{0.42\hsize}{!}
{\includegraphics[width=0.47\linewidth]{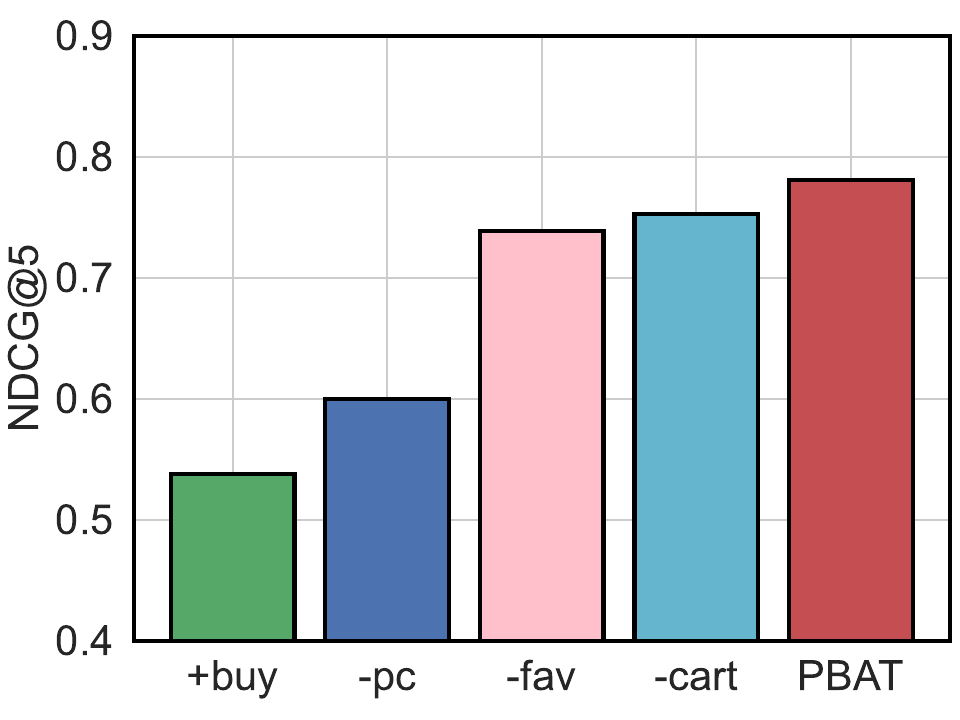}}}
 \vspace{-0.2 in}
\caption{Data ablation studies on auxiliary behaviors.}
 \vspace{-0.25 in} 
\label{fig:ablation2}
\end{figure}

\subsection{Ablation Study (RQ2)}
To evaluate the rationality of two main modules, we conduct ablation studies on the following variants:
(1) \textbf{\textit{w/o PBPG}} removes the personalized behavior pattern generator module, which means removing personalized pattern learning but keeping dynamic representation encoding for attention layers.
    %
%
(2) \textbf{\textit{w/o BACE}} removes the behavior-aware collaboration extractor module, which means adopting self-attention layers in \cite{sun2019bert4rec} without performing fused behavior-aware attention.
%
To study effectivenesses of each key component in the behavior-aware collaboration extractor, we further consider:
(3) \textbf{\textit{w/o fob}} removes behavioral factor in PB-Fusion, and replaces TriSAGP method with SAGP method between item and position representations.
(4) \textbf{\textit{w/o fop}} removes temporal factor in PB-Fusion.
(5) \textbf{\textit{w/o fobp}} considers neither behavioral nor temporal factor in PB-Fusion, and removes the TriSAGP process totally.

We present the results in Table \ref{tab:ablation}, from which we can observe that:
(1) \modelname~outperforms both \textit{w/o PBPG} and \textit{w/o BACE} variants, indicating the effectiveness of two modules, i.e., personalized behavior pattern generator and behavior-aware collaboration extractor, respectively.
    Comparing \textit{w/o PBPG} and \textit{w/o BACE}, \textit{w/o BACE} performs worse on three datasets, which suggests that the fused behvaior-aware attention mechanism we employ in the attention layer contributes more than personalized pattern learning to \modelname.
    Besides, the performance of \textit{w/o BACE} is comparable with some strong MBSR baselines, e.g., DMT.  
    This phenomenon demonstrates that utilizing distribution embeddings for encoding entities or relations in the transformer-based model extracts more robust and discriminative representations, and thus improves the recommendation accuracy of the backbone.
(2) The performance gap between \modelname~and variant \textit{w/o fobp} proves that incorporating behavioral and temporal impacts into the collaborative transitions significantly promotes sequential modeling. 
    \modelname~also performs better than both variants \textit{w/o fob} and \textit{w/o fop}, showing that it brings performance degradation when removing either influence factor, i.e., behavior or position, in TriSAGP process. 
    Further, \textit{w/o fop} beats \textit{w/o fob} on all metrics, from which we conclude the behavior transition pattern has a more dominant impact than temporal intervals on depicting multifaceted collaborations for multi-behavior sequences.

\begin{figure}[t]
 \begin{center}
 \includegraphics[width=\columnwidth]{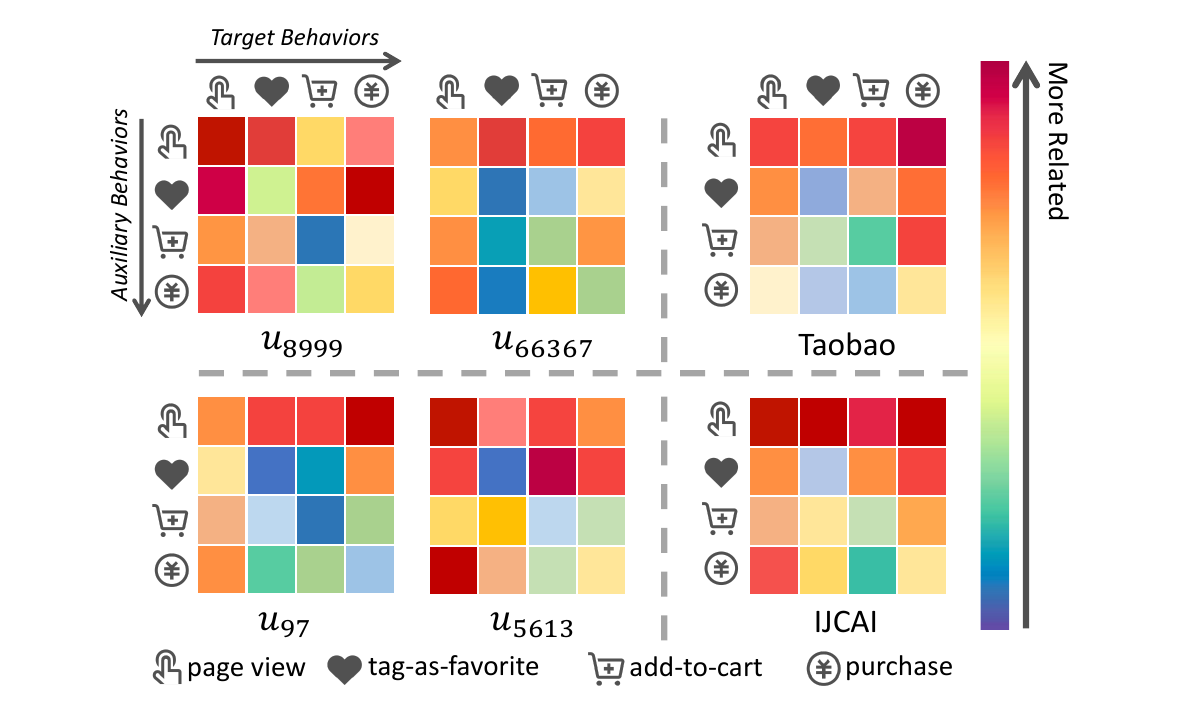}
 \vspace{-0.3 in}
 \caption{Case studies with multi-behavior patterns.
 }
 \vspace{-0.25 in} 
 \label{fig:case-study}
 \end{center}
\end{figure}
\subsection{Effect of Auxiliary Behaviors (RQ3)}
We further perform data ablation studies to validate the contributions of different types of auxiliary behaviors.
We investigate four variants on datasets Taobao and IJCAI, where  \textit{-pv}, \textit{-fav}, \textit{-cart} represent our model \modelname~without the incorporation of \textit{page view}, \textit{tag-as-favorite}, \textit{add-to-cart} behavior, respectively.
And \textit{+buy} denotes the variant which only utilizes the target behavior.
We present the results in Figure \ref{fig:ablation2}.
The results show that: (1) \modelname~performs the best, compared with all variants, which demonstrates the effectiveness of leveraging comprehensive multi-behavior patterns in user preference modeling.
(2) The performance declines when removing any type of auxiliary behaviors, and the descent of removing \textit{page view} is most significant.
This phenomenon indicates that \textit{page view} is a much more informative factor in behavior pattern generalization, and has a stronger correlation with \textit{purchase} behavior.

\subsection{Model Interpretation Study (RQ4)}
We further perform case studies with sampled user examples to show interpretation capability of \modelname~for personalized multi-behavior pattern learning.
In our learning framework, all behavior types could be regarded as the auxiliary behavior or the target behavior.
Particularly, we present behavior patterns that \modelname~ captures on dataset Taobao and IJCAI in Figure \ref{fig:case-study}.
We generate a 4×4 dependency matrix corresponding to correlation coefficients between each auxiliary behavior and target behavior in the attention layer.
Here, we sample four user-specific cases ($u_{8999}$ and $u_{66367}$ for Taobao, $u_{97}$ and $u_{5613}$ for IJCAI), and we show the overall behavior-relation matrix without taking user representation into consideration on the right side.
From it, we can observe that: (1) Two datasets exhibit different general features on behavior patterns, since they cover different user groups and item sets.
(2) For both datasets, \textit{page view} contributes the most to predicting other types of behaviors.
It is potentially because \textit{page view} is the most frequent type of behaviors for users in these datasets.
%
%
(3) Users show distinguished behavior patterns due to their personal habits and characteristics.
For example, $u_{8999}$ was used to tag items as favorite before he finally bought them.
But for $u_{66367}$, add-to-cart is more informative than tag-as-favorite when inferring a purchase behavior.

\subsection{Parameter analysis (RQ5)}
\begin{figure} 
\centering
\subfigure[$D$-Taobao]{\label{fig:subfig:p-a}
\resizebox{0.42\hsize}{!}
{\includegraphics[width=0.47\linewidth]{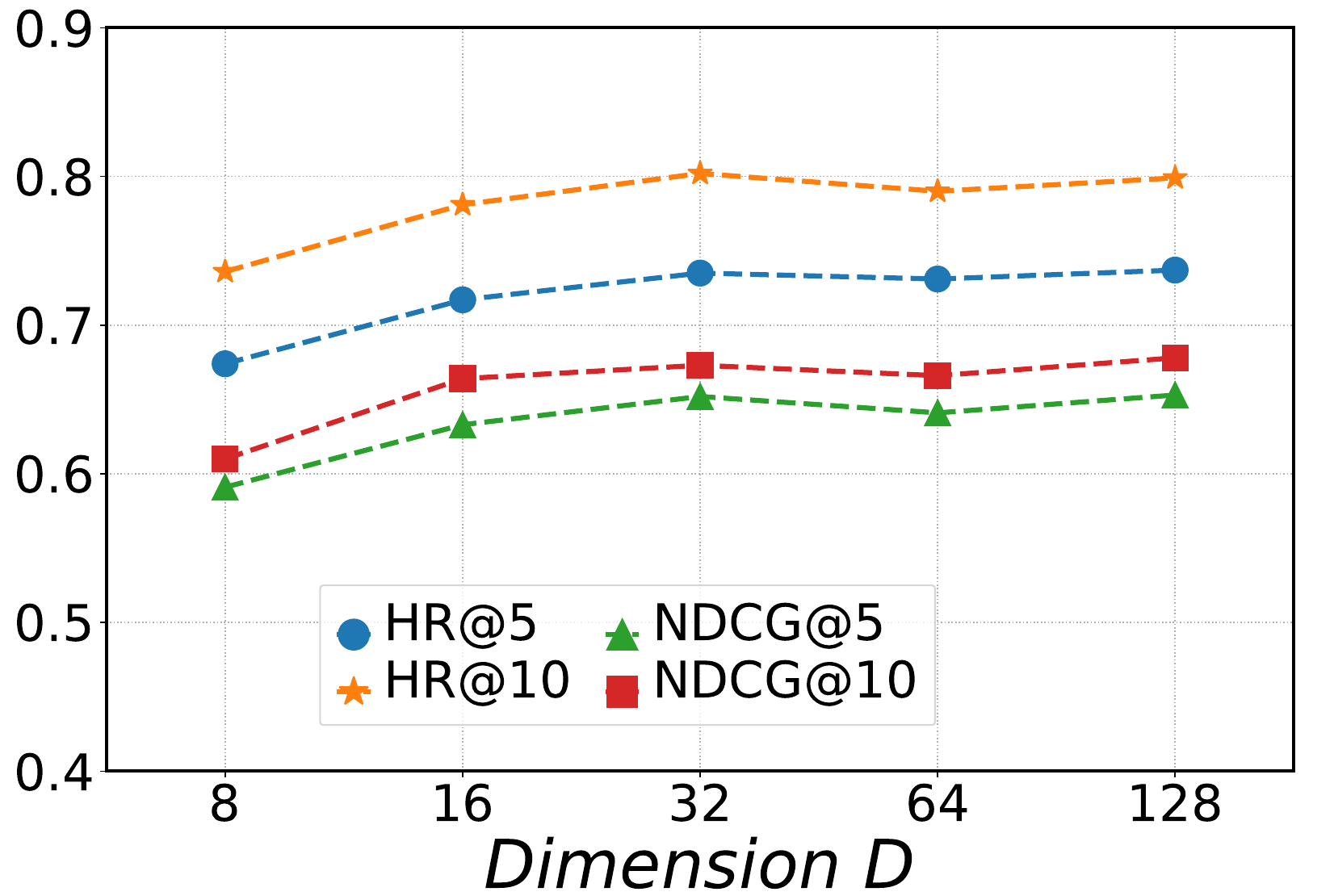}}}
\hspace{0.01\linewidth}
\subfigure[$D$-IJCAI]{\label{fig:subfig:p-b}
\resizebox{0.42\hsize}{!}
{\includegraphics[width=0.47\linewidth]{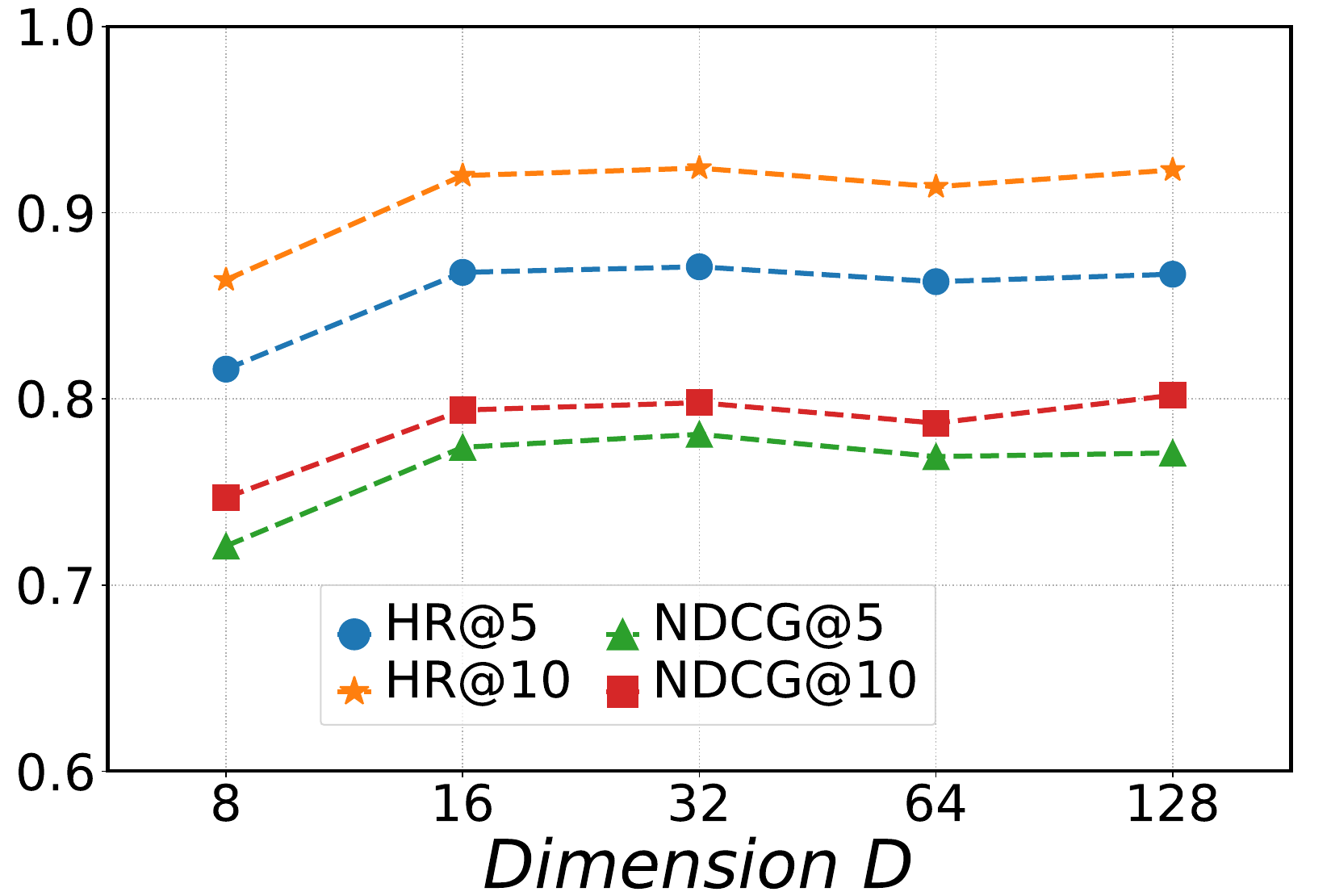}}}\\
\subfigure[$N$-Taobao]{\label{fig:subfig:p-c}
\resizebox{0.42\hsize}{!}
{\includegraphics[width=0.47\linewidth]{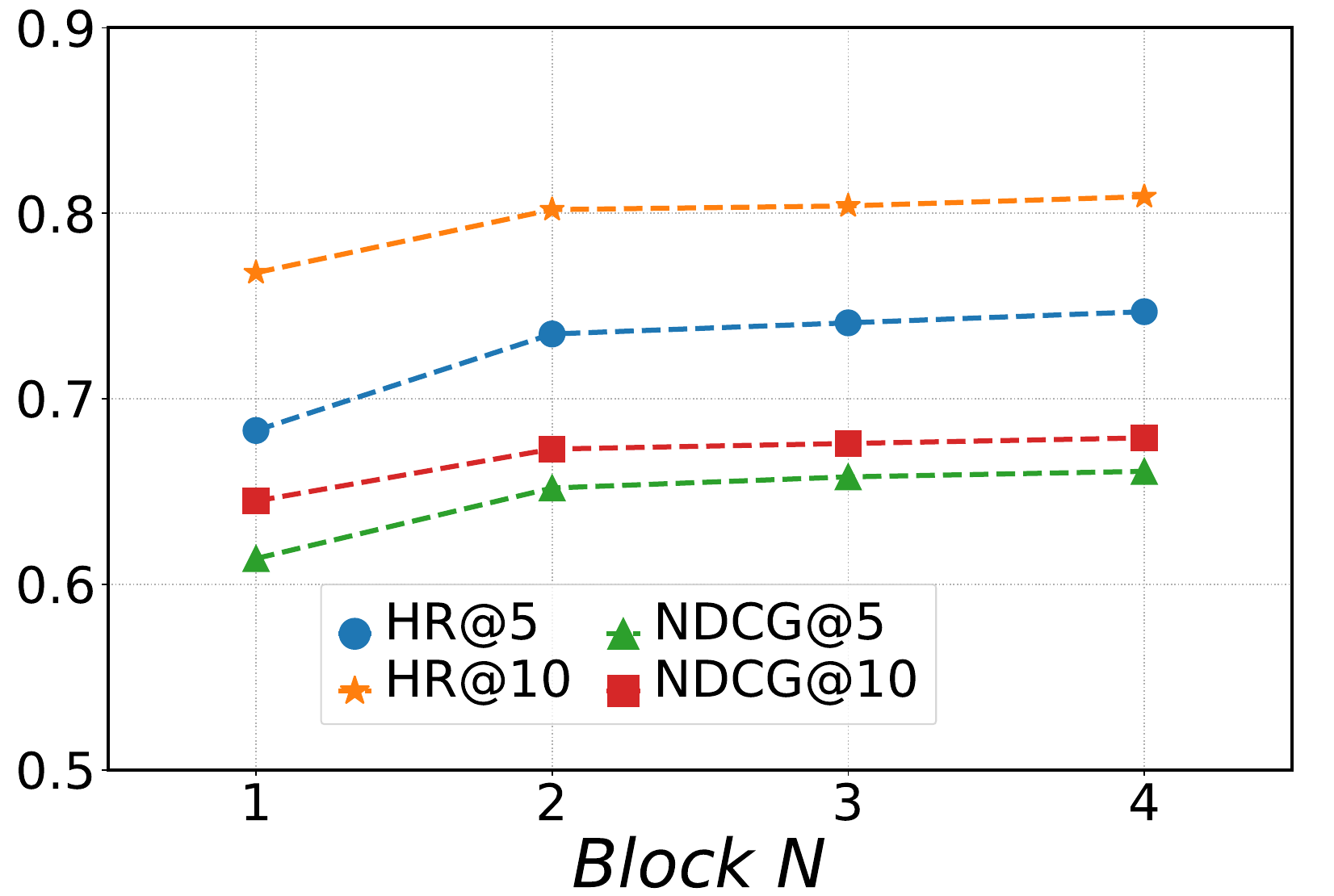}}}
\hspace{0.01\linewidth}
\subfigure[$N$-IJCAI]{\label{fig:subfig:p-d}
\resizebox{0.42\hsize}{!}
{\includegraphics[width=0.47\linewidth]{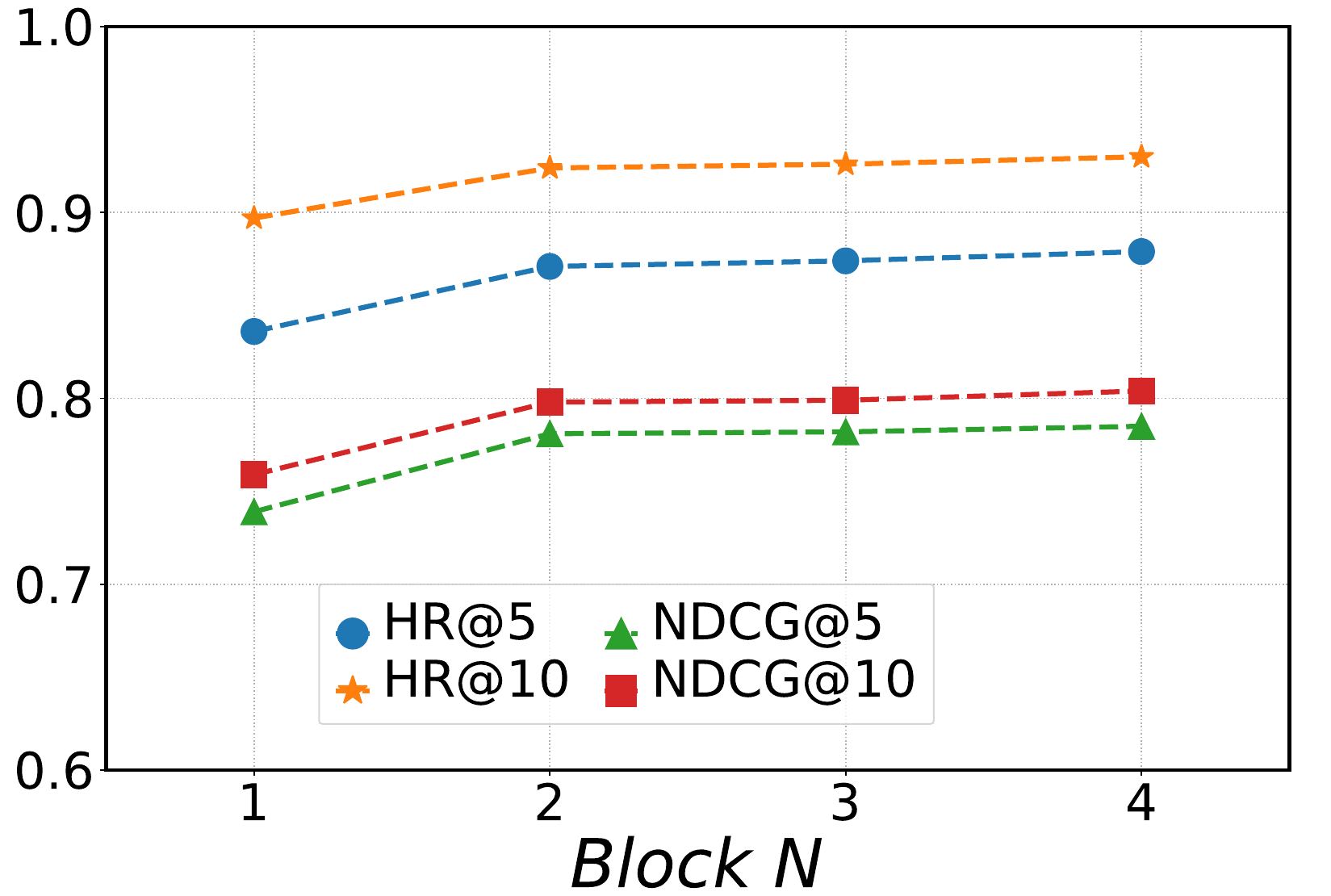}}}
\vspace{-0.2 in}
\caption{Effects of $D$ and $N$ on \modelname.}
\label{fig:parameter}
\vspace{-0.25 in}
\end{figure}
We now study the effects of hyper-parameters on model performance, including embedding dimension $D$, and self-attention block number $N$.
We first study the impact of embedding dimension on \modelname~by varying $D$ in $\{8,16,32,64,128\}$ and reporting results on Taobao and IJCAI in Figure \ref{fig:parameter}(a)-(b).
The results on Yelp are in Appendix E.
From the results, we can observe that the performance gradually improves when $D$ increases and keeps a fairly stable level after it reaches a certain threshold.
It demonstrates that representations with larger dimensions are more informative and accurate for portraying entity and distribution features.
Considering the computational and time cost for training, we choose $D=32$ for Taobao, $D=16$ for IJCAI and Yelp, respectively.
Then we perform experiments by varying the number of blocks $N$ in $\{1,2,3,4\}$.
We present the results on two datasets in Figure \ref{fig:parameter}(c)-(d).
The curves show that the recommendation accuracy boosts first and then levels off on both datasets.
When the self-attention network structure is relatively shallow, it cannot fully extract multi-behavior sequential dependencies.
But when we stack too many blocks, the performance gain brought by the perceptron is limited.
In this case, we choose $N=2$ for all datasets to realize stable and effective model training.

\section{Conclusion}
In this paper, we propose a transformer-based model \modelname~, which explores personalized multi-behavior patterns and multifaceted time-evolving collaborations for MBSR problem.
The framework consists of two modules, i.e., personalized behavior pattern generator and behavior-aware collaboration extractor.
In the pattern generator, we employ dynamic representation encoding and personalized pattern learning to capture distinguishing behavior patterns.
In the collaboration extractor, we extract comprehensive correlations among behavioral and temporal factors, further promoting the modeling of user preferences.
Extensive experiments conducted on public datasets illustrate the effectiveness of \modelname.

\begin{acks}
This work was supported in part by the National Key
R\&D Program of China (No. 2022YFF0902704) and the Ten
Thousand Talents Program of Zhejiang Province for Leading
Experts (No. 2021R52001).
\end{acks}

\bibliographystyle{ACM-Reference-Format}
\balance
\bibliography{sample-base}


\appendix
\section{Notations}\label{sec:note}
We summarize the main notations used in this paper in Table \ref{tab:notation}.

\begin{table}[htbp]
\small
  \centering
  \caption{Notation table.}
    \begin{tabular}{cc}
    \toprule
    \textbf{Notation} & \textbf{Meaning} \\
    \hline
    $\mathcal{U},\mathcal{V},\mathcal{B}$ & set of user, item, and behavior\\
    $\mathcal{S}_u$ & multi-behavior sequence of user $u$\\
    $D$ & dimension of embedding\\
    $L$ & multi-behavior sequence length \\
    $N$ & self-attention block number\\
    $\textbf{H}^\mu, \textbf{H}^\sigma$ & mean and covariance on item distribution\\
    $\textbf{E}^\mu, \textbf{E}^\sigma$ & mean and covariance on user distribution\\
    $\textbf{B}^\mu, \textbf{B}^\sigma$ & mean and covariance on behavior distribution\\
    $\textbf{P}^\mu, \textbf{P}^\sigma$ & mean and covariance on position distribution\\
    $\textbf{R}^\mu, \textbf{R}^\sigma$ & mean and covariance on behavior-relation distribution\\
    $\mathcal{N}(\mu^{pt}_{u,i}, \sigma^{pt}_{u,i})$ & personalized pattern of user $u$ on behavior $i$\\
    $\mathcal{N}(\mu^{ip}_{s,t}, \sigma^{ip}_{s,t})$ & behavioral collaboration impact factor between $v_s$ and $v_t$\\
    $\mu^{Q},\mu^{K},\mu^{V}$ & mean on initialized queries, keys, and values\\
    $\sigma^{Q},\sigma^{K},\sigma^{V}$ & covariance on initialized queries, keys, and values\\
    $\mu^{Q'},\mu^{K'}$ & mean on queries and keys after TriSAGP\\
    $\sigma^{Q'},\sigma^{K'}$ & covariance on queries and keys after TriSAGP\\
    $\alpha_{s,t}$ & attention score of $v_s$ and $v_t$\\
    $X_\mu, X_\sigma$ & outputs of self-attention blocks \\
    \bottomrule
    \end{tabular}%
  \label{tab:notation}%
\end{table}%

\section{Behavior-Aware Collaboration Extractor}\label{sec:algo}
We represent details of behavior-aware collaboration extractor in Algorithm \ref{algo}.
%
%
We divide it into five steps, i.e., initialize queries, keys, values (line 1-4), get behavioral collaboration impact factor (line 6-9), position-enhanced behavior-aware fusion (line 10-12), get attention scores (line 13-14), and attentive aggregation (line 16-20).

\section{Design of Feed-forward layer and Self-attention block}\label{sec:ffl-sab}
We design feed-forward layers (FFL) in our transformer-based model to integrate sequential context.
To capture behavioral semantics, we propose behavior-specific FFL which utilize distinct multilayer perceptrons for different behavior types:
\begin{equation}
    f^{\mu}_t = FFL(x^\mu_t) = ELU(x^\mu_t\textbf{W}_\mu^{BT1}+b_\mu^{BT1})\textbf{W}_\mu^{BT2}+b_\mu^{BT2}. 
\nonumber
\end{equation}
Here, $x^\mu_t$ is the mean embedding output at position $t$ of one attention layer.
$\textbf{W}_\mu^{BT1}, \textbf{W}_\mu^{BT2}, b_\mu^{BT1}, b_\mu^{BT2}$ are parameters of behavior-specific multilayer perceptrons, where $BT$ indicates the behavior type at $t$-th position.
As previous studies \cite{kang2018self,sun2019bert4rec,yuan2022multi}, it is beneficial to stack transformer layers for extracting more complex item transitions. 
Specifically, we design the self-attention blocks (SAB) by adopting the similar formulas of residual connection, layer normalization, and dropout layers from BERT4REC \cite{sun2019bert4rec}.
The output of mean embeddings generated by the $n$-th block is defined as:
\begin{equation}
\begin{aligned}
    F_\mu&=\left(f_1^\mu,f_2^\mu,\ldots,f_L^\mu\right), X_\mu^{(n)} = SAB\left(F^{(n-1)}_\mu\right), \quad \forall n>1.
\end{aligned}
\nonumber
\end{equation}
Similarly, the outputs of covariance can be generated as $X_\sigma^{(n)}$.
%
\begin{algorithm}[H]
\caption{Behavior-Aware Collaboration Extraction}\label{algo}
\begin{algorithmic}[1]
    \Statex \textbf{Input:} ItemEmb $\mathcal{N}(\mu^v,\sigma^v)$, BehaviorEmb $\mathcal{N}(\mu^b,\sigma^b)$, PositionEmb $\mathcal{N}(\mu^p,\sigma^p)$, PatternEmb $\mathcal{N}(\mu^{pt}_u,\sigma^{pt}_u)$, BehaviorRelation $[R_\mu,R_\sigma]$
    \Statex \textbf{Output: } Attention layer outputs from one head $X_\mu$, $X_\sigma$ 
    \State \textbf{Step1: Initialize Queries, Keys, Values}
    \State $\mu^Q,\sigma^Q \gets f_Q(\mu^v,\sigma^v,\mu^b,\sigma^b)$
    \State $\mu^K,\sigma^K \gets f_K(\mu^v,\sigma^v,\mu^b,\sigma^b)$
    \State $\mu^V,\sigma^V \gets f_V\mu^v,\sigma^v,\mu^b,\sigma^b)$
    
    \For{$s=1$ to $L,\ t=1$ to $L$}
    \State \textbf{Step2: Get Behavioral Collaboration Impact Factor}
    \State $\mu^{ptc}_u,\sigma^{ptc}_u \gets f_{pt}(\mu^{pt}_u,\sigma^{pt}_u)$
    \State $m^u_{s,t} \gets WassDis(\mathcal{N} (\mu^{ptc}_{u,s},\sigma^{ptc}_{u,s}),\mathcal{N}(\mu^{ptc}_{u,t},\sigma^{ptc}_{u,t}))$
    \State $\operatorname{Impact\ Factor\ \textbf{ip}}\gets \mathcal{N}(m^u_{s,t}\boldsymbol{r}^\mu_{b_sb_t},m^u_{s,t}\boldsymbol{r}^\sigma_{b_sb_t})$
    \State \textbf{Step3: Position-Enhanced Behavior-Aware Fusion}
    \State $\mathcal{N}(\mu^{K'}_s,\sigma^{K'}_s)\gets TriSAGP([\mu^K_s,\sigma^K_s],[\mu^{ip}_{s,t},\sigma^{ip}_{s,t}],[\mu^{p}_{s},\sigma^{p}_{s}])$
    \State $\mathcal{N}(\mu^{Q'}_t,\sigma^{Q'}_t)\gets TriSAGP([\mu^Q_t,\sigma^Q_t],[\mu^{ip}_{s,t},\sigma^{ip}_{s,t}],[\mu^{p}_{t},\sigma^{p}_{t}])$
    \State \textbf{Step4: Get Attention Scores}
    \State $\alpha_{s,t}\gets-WassDis(\mathcal{N}(\mu^{K'}_s,\sigma^{K'}_s),\mathcal{N}(\mu^{Q'}_t,\sigma^{Q'}_t))$
    \EndFor
    \State \textbf{Step5: Attentive Aggregation}
    \For{$t=1$ to $L$}
        \State $x^\mu_t \gets \sum^L_{j=1}\frac{\alpha_{j,t}}{\sum^L_{i=1}\alpha_{i,t}}\mu^V_j$
        \State  $x^\sigma_t \gets \sum^L_{j=1}{\left(\frac{\alpha_{j,t}}{\sum^L_{i=1}\alpha_{i,t}}\right)}^2\sigma^V_j$
    \EndFor
    \State $X_\mu \gets (x^\mu_1, x^\mu_2,\ldots,x^\mu_L)$
     \State $X_\sigma \gets (x^\sigma_1, x^\sigma_2,\ldots,x^\sigma_L )$
\end{algorithmic}
\end{algorithm}

\section{DATASETS DESCRIPTION}\label{sec:dataset}
We conduct extensive experiments on three popularly used real-
world datasets, i.e., \textbf{Taobao}, \textbf{IJCAI}, and \textbf{Yelp}.
As previously mentioned, both of \textbf{Taobao} and \textbf{IJCAI} contain four types of behavior, i.e., \textit{page view}, \textit{tag-as-favorite}, \textit{add-to-cart}, and \textit{purchase}.
As for \textbf{Yelp}, we differentiate the rating data from 1 to 5 with 0.5 as increment, and generate three types of behavior, i.e., \textit{like} (rating $\geq 4$), \textit{neutral} (rating $>2$ and $<4$), and \textit{dislike} (rating $\leq 2$).
Besides, we also include the behavior \textit{tip} which already existed in the original dataset.
We show the statistics of three datasets in Table 
\ref{tab:dataset}.
\begin{table}[H]
\small
  \centering
  \caption{Statistics of the used datasets.}
    \begin{tabular}{ccccc}
    \toprule
    
    \textbf{Datasets}  & \textbf{\#User} & \textbf{\#Items} &  \textbf{\#Interactions} & \textbf{Behavior Types}\\
    \hline 
    Taobao & 147,894 & 99,037 &  7,658,926 &  [pv, fav, cart, purchase]\\
    IJCAI & 423,423 & 874,328 & 36,222,123 & [pv, fav, cart, purchase]\\
    Yelp & 19,800 & 22,734 &1,400,002 & [tip, dislike, neutral, like]\\
    
\bottomrule
    \end{tabular}%
  \label{tab:dataset}%
\end{table}%

\section{Parameter analysis on dataset Yelp}\label{sec:yelp}

\begin{figure}[H]
\centering
\subfigure[$D$-Yelp]{\label{fig:subfig:yelp-a}
\includegraphics[width=0.47\linewidth]{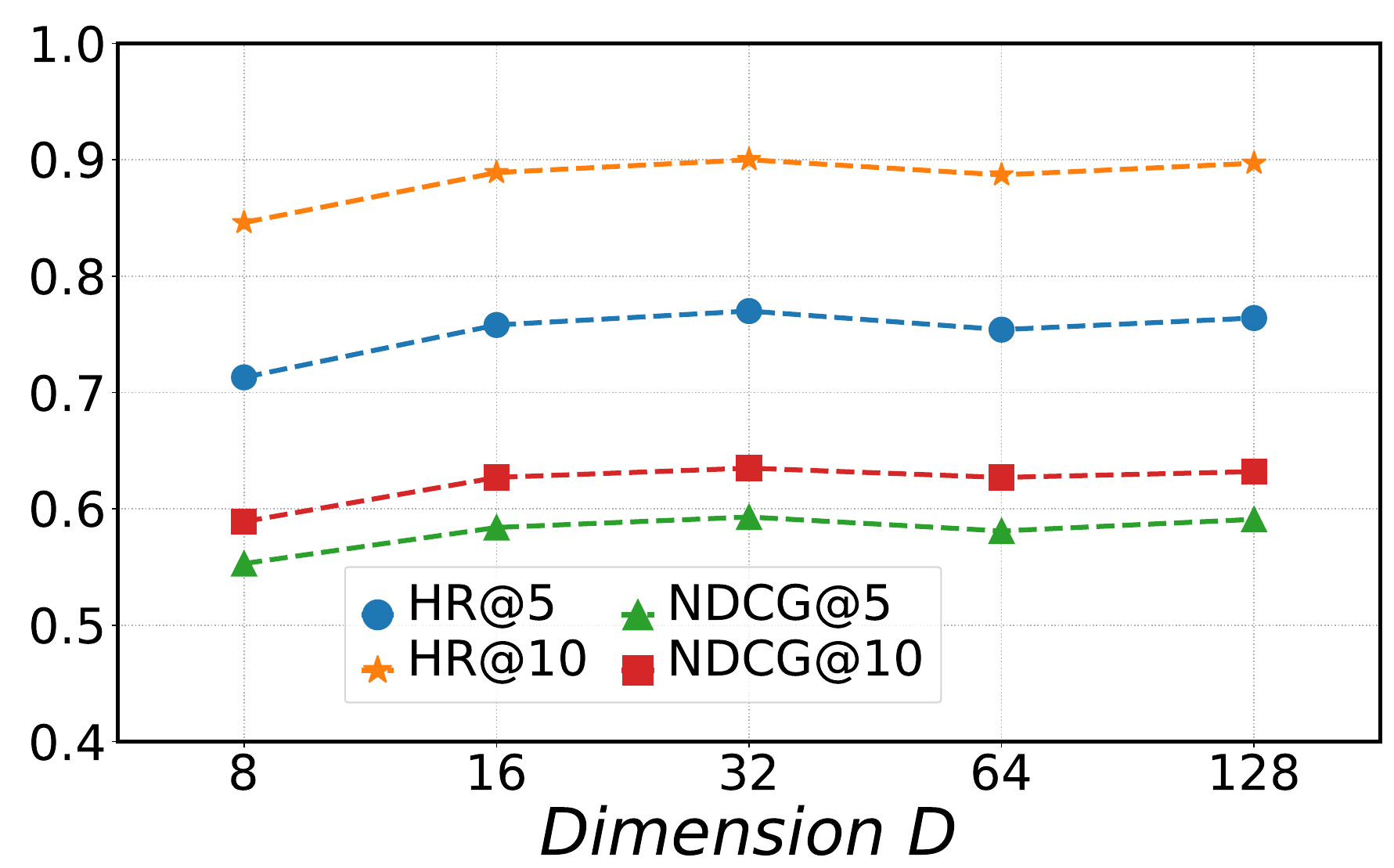}}
\hspace{0.01\linewidth}
\subfigure[$N$-Yelp]{\label{fig:subfig:yelp-b}
\includegraphics[width=0.47\linewidth]{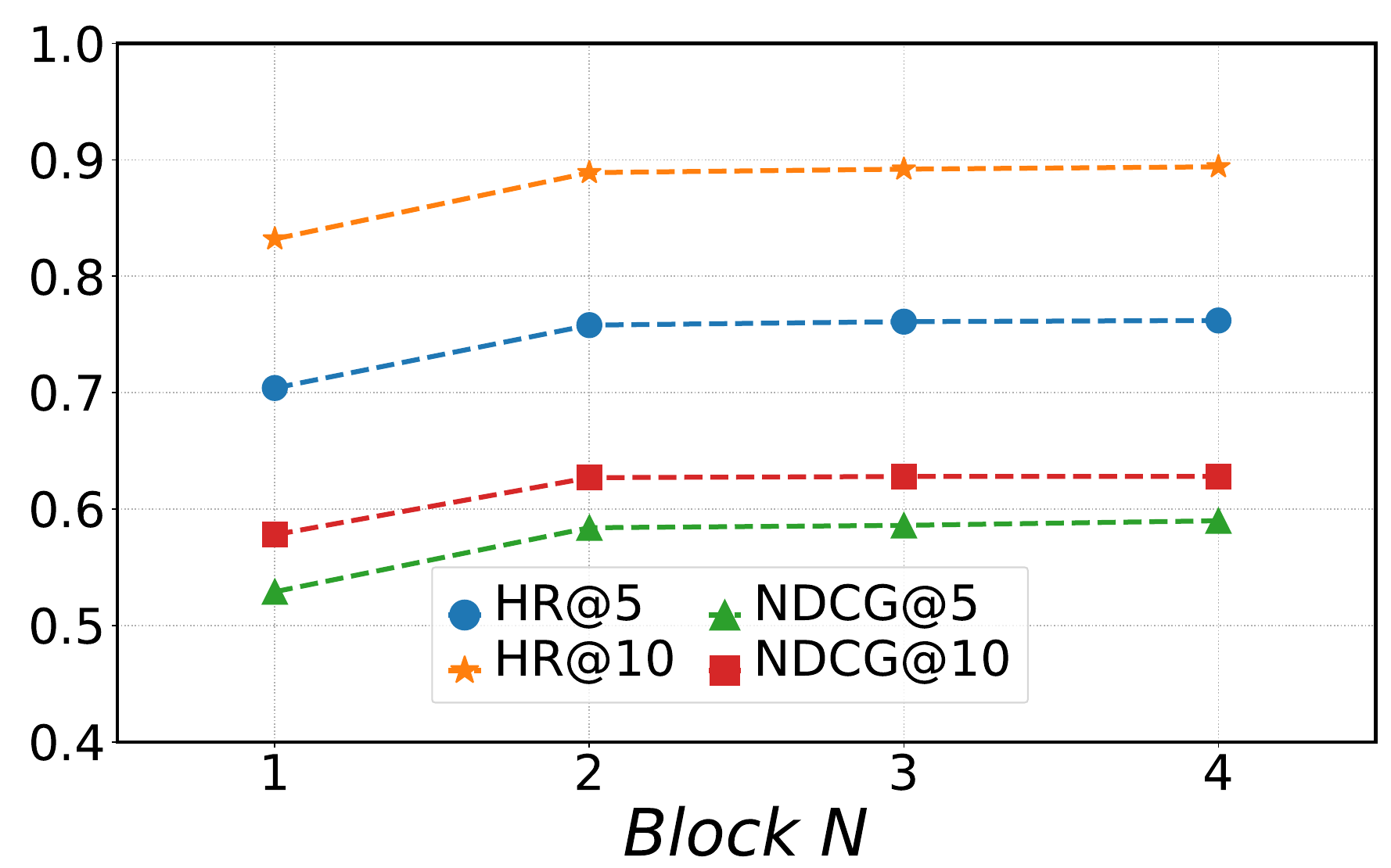}}
\caption{Effects of $D$ and $N$ on \modelname.}
\label{fig:parameter-yelp}
\end{figure}

We present effects of embedding dimension $D$, self-attention block
number $N$ on dataset \textbf{Yelp}.
The results are shown in Fig \ref{fig:subfig:yelp-a}-\ref{fig:subfig:yelp-b}.
We conclude that: (1) the performance of \modelname~boosts when $D$ increases from 8 to 16.
But after $D$ reaches 16, the metrics no longer improves significantly.
(2) As the number of self-attention blocks $N$ increases, the recommendation accuracy improves first and then levels off.
In practice, we set $D=16$ and $N=2$ for \textbf{Yelp}.

\end{document}